\documentclass[12pt]{article}
\usepackage{amsmath,amssymb,epsfig}
\usepackage{cite}
\usepackage{hyperref}
\usepackage{here}

\usepackage{color}
\input{colordvi.tex}
\def\unit{{\relax{\rm 1\kern-.26em I}}}

\def\beq{\begin{equation}}
\def\eeq{\end{equation}}
\newcommand{\bea}{\begin{eqnarray}\begin{aligned}}
\newcommand{\eea}{\end{aligned}\end{eqnarray}}

\makeatletter

\renewcommand\section{\@startsection {section}{1}{\z@}%
                                   {-3.5ex \@plus -1ex \@minus -.2ex}%
                                   {2.3ex \@plus.2ex}%
                                   {\normalfont\large\bfseries}}

\renewcommand\subsection{\@startsection{subsection}{2}{\z@}%
                                     {-3.25ex\@plus -1ex \@minus -.2ex}%
                                     {1.5ex \@plus .2ex}%
                                     {\normalfont\normalsize\bfseries}}

\makeatother

\begin{document}

\baselineskip=18pt  
\numberwithin{equation}{section}  
\allowdisplaybreaks  



%
%


\thispagestyle{empty}

\vspace*{-2cm}
\begin{flushright}
\end{flushright}

\begin{flushright}

\end{flushright}

\begin{center}

\vspace{-0.2in}
{\bf \LARGE
Strange Jet Tagging } 
\vspace*{0.45in}

{ \large
Yuichiro Nakai$^1$, David Shih$^{2,3,4}$ and Scott Thomas$^2$} \\
\vspace*{0.2in}

{\it $^1$ Tsung-Dao Lee Institute and School of Physics and Astronomy, \\
Shanghai Jiao Tong University, 800 Dongchuan Road, Shanghai 200240, China}\\
\vspace*{0.1in}

{\it $^2$New High Energy Theory Center, Department of Physics and Astronomy, \\
   Rutgers University, Piscataway, NJ 08854, USA}\\
\vspace*{0.1in}

{\it $^3$Theory Group, Lawrence Berkeley National Laboratory,\\ Berkeley, CA 94720, USA}\\
\vspace*{0.1in}

{\it $^4$Berkeley Center for Theoretical Physics, \\University of California, Berkeley, CA 94720, USA
}\\

\vspace*{0.2cm}

\end{center}

\vspace{1cm} \centerline{\bf Abstract} \vspace*{0.5cm}

\baselineskip=16pt Tagging jets of strongly interacting particles 
initiated by energetic strange quarks is one of the few largely 
unexplored Standard Model object classification problems remaining 
in high energy collider physics. 
In this paper we investigate the purest version of this classification problem 
in the form of distinguishing strange-quark jets from down-quark jets. 
Our strategy relies on the fact that 
a strange-quark jet contains on average a higher ratio of  
neutral kaon energy  
to neutral pion energy than does a down-quark jet.  
Long-lived neutral kaons deposit energy mainly in the hadronic calorimeter 
of a high energy detector, while neutral pions decay promptly to photons 
that deposit energy mainly in the electromagnetic calorimeter. 
In addition, short-lived neutral kaons that decay in flight 
to charged pion pairs can be identified as a secondary vertex 
in the inner tracking system. 
Using these handles we study different approaches to distinguishing 
strange-quark from down-quark jets, 
including single variable cut-based methods, 
a boosted decision tree (BDT) with a small number of simple variables, 
and a deep learning convolutional neural network (CNN) architecture with jet images.
We show that modest gains are possible from the CNN compared with 
the BDT or a single variable. 
Starting from jet samples with only strange-quark and down-quark jets, 
the CNN algorithm can improve the strange to down ratio 
by a factor of roughly 2 for strange tagging efficiencies below 0.2, and 
by a factor of 2.5 for strange tagging efficiencies near 0.02.

\newpage
\setcounter{page}{1} 

\baselineskip=18pt

\section{Introduction\label{sec:intro}}

Understanding jets, collimated collections 
of hadrons arising from strongly interacting particles produced in high-energy collisions,
is a key ingredient of physics measurements and new physics searches at 
a high energy collider such as the Large Hadron Collider (LHC).
One of the goals of jet reconstruction in a modern high energy collider detector 
is to use the observable properties of the jet to determine on average the 
identity of the strongly interacting particle that initiated the jet 
before QCD showering and hadronization. 

Various methods for jet tagging have been developed over the years. 
Gluon jets tend to have more constituents with more uniform energy 
fragmentation and wider spread, as compared to light-quark jets, owing 
to their larger color factor (see e.g. Ref.~\cite{Asquith:2018igt} and references therein). 
For recent papers that apply machine learning to the problem of quark/gluon tagging, 
see Refs.~\cite{Komiske:2016rsd,Cheng:2017rdo,Luo:2017ncs,Komiske:2018oaa,Fraser:2018ieu,Metodiev:2017vrx,Kasieczka:2018lwf}.
Bottom- and charm-quark tagging can be achieved by looking for 
displaced charged track vertices \cite{Aad:2015ydr,Sirunyan:2017ezt}.
For up- versus down-quark tagging, the $p_T$-weighted track charge can offer some level of 
discrimination 
 \cite{Field:1977fa,Krohn:2012fg,Waalewijn:2012sv,Fraser:2018ieu}.
Finally, if a top-quark is boosted, its decay products 
tend to be collimated into a large-radius jet that can be 
distinguished from quark/gluon jets by using jet mass or jet 
substructure variables such as N-subjettiness \cite{Thaler:2010tr} 
(for a review of boosted top tagging and original references, see Ref.~\cite{Plehn:2011tg}).
Recently, deep learning has been applied to the problem of 
boosted top-quark tagging with great success~\cite{ANNTop,CanadaTop,DeepTop,RuDeTop,RecursiveTop,CanadaLSTM,LoLa,LBN,NSubTop,LDA,EFN,ParticleNet,Moreno:2019neq,CMSTop,ATLASTop,Landscape}.

The last largely unexplored type of jet tagging is for strange-quark jets.  
In this work, we investigate 
the possibilities for strange-quark jet tagging using observables 
that are available in the current generation of general purpose 
high energy collider detectors, such as the CMS and ATLAS detectors at the LHC. 
Strange-quark jet tagging at future $e^+e^-$ colliders has also recently 
been studied using Monte Carlo truth-level information \cite{Duarte-Campderros:2018ouv}.
Deployment of strange-quark jet tagging in the analysis of actual data from a high energy collider 
would of necessity be within the context of distinguishing among all types of prompt jets.  
In this demonstration-level 
work we restrict attention to the more restrictive problem of distinguishing strange- and down-quark 
jets.  
Jets initiated by these two types of quarks are most alike, and so the 
discrimination problem is most challenging.  

Strange- and down-quarks have identical QCD and electromagnetic interactions, and 
so effectively only differ in hadronization and subsequent decay processes. 
As detailed below, the key difference at the hadronization level is that strange-quark jets 
have a larger momentum weighted fraction of kaon mesons than do down-quark jets. 
Long-lived neutral kaons deposit energy primarily in the hadronic calorimeter (HCAL) 
of a general purpose collider detector.  
In contrast, neutral pions, that are more common in down-quark jets, decay promptly 
to pairs of photons that deposit energy in the electromagnetic calorimeter (ECAL). 
So on average, a strange-quark jet has a somewhat larger neutral hadronic energy fraction, 
while a down-quark jet has a larger neutral electromagnetic energy fraction. 
This difference can provide a handle to on average distinguish strange- and down-quark jets. 
In addition, the momentum of 
short-lived neutral kaons that decay-in-flight to charged pion pairs within the inner tracking 
region can be reconstructed \cite{Khachatryan:2010pw,Aad:2011hd} 
and this provides another handle to distinguish these types of jets.  

We explore a number of strange- versus down-quark jet tagging algorithms based on 
observables available in the current generation of general purpose high energy collider detectors.  
These range from single whole-jet variables, 
to Boosted Decision Trees (BDTs) with a few whole-jet variables, 
to Convolutional Neutral Networks (CNNs) with jet images based on inputs 
from detector sub-systems.  
The CNNs provide modest gains in discrimination, with the best one making use 
of all of the available handles mentioned above.  

Jet tagging algorithms have many applications in high energy collider physics. 
Sub-dividing events into categories based on tagging algorithm outputs 
can be used to isolate sought out signals from backgrounds, thereby 
increasing the sensitivity of new physics searches or improving the precision 
of measurements.  
Adding strange-quark jet tagging could improve existing or future searches, or even make 
new types of measurements possible. 
In conjunction with bottom- and charm-quark tagging with displaced vertices,  
strange-quark jet tagging could help 
reduce combinatoric confusion in the reconstruction of resolved hadronic top-quark decays. 
Ultimately this might also make possible a direct measurement of the 
ratio of CKM matrix elements $| V_{cs} / V_{cd}|$ in hadronic $W$-boson decays. 

In the next section, the event generation and detector simulation used in the studies 
here are presented. 
In section 3, the differences between strange- and down-quark jets are detailed, 
both at the truth-level as well as at the level of observables available to a general purpose 
collider detector. 
In section 4, the strange- versus down-quark tagging algorithms developed here are defined. 
In section 5, the quantitative performance of these tagging algorithms is assessed. 
A general discussion of improvements offered by tagging algorithms is presented 
in section 6, along with a possible application of strange-quark jet tagging to hadronic 
top-quark reconstruction. 
Discussion of future directions, including possible improvements to strange-quark jet 
tagging in  future collider detectors, are given in the final section.


\section{Event Generation and Detector Simulation}
\label{sec:methods}

In order to investigate the discrimination of strange- and down-quark jets, 
Monte Carlo samples of di-jet events 
 for 13 TeV proton-proton collisions are simulated using 
\textsc{MadGraph5} version 2.6.0 \cite{Alwall:2011uj} for the hard scattering processes, along with 
\textsc{Pythia8} \cite{Sjostrand:2014zea}
for showering and hadronization. 
Two sets of 
one million scattering events through an intermediate $Z$-boson, 
$pp \to Z \to s \bar{s}$ and $pp \to Z \to d \bar{d}$, 
are simulated to provide samples of intermediate momentum jets in the final state. 
Another two sets of one million scattering events through QCD processes, 
$p p \to s \bar{s}$ and $pp\to  d \bar{d}$, with $p_{T}$ of the leading hard scattering quark 
in the final state greater than 200 GeV,
 are simulated to provide samples of higher 
momentum jets. 
In all cases, the pseudo-rapidity of the hard scattering final state quarks are required 
to be $| \eta | < 0.05$. 
The restriction to very close to lab-frame central scattering ensures that the kinematics of the hard scattering 
events, as well as the response of the detector simulation discussed below, 
are uniform across an entire sample. 
This allows for an unbiased comparison among the samples. 
The intermediate $Z$-boson and QCD samples are referred to below as the $p_T = 45$ GeV
and $p_T > 200$ GeV samples, respectively. 
We also generated similar samples with \textsc{Herwig} \cite{Bahr:2008pv,Bellm:2015jjp}
with almost no differences, and only present results from the \textsc{Pythia8} samples in this paper.

The \textsc{Pythia8} output of hadronized particles for each event in the samples 
is passed to the \textsc{Delphes} fast detector simulator version 3.4.1 \cite{deFavereau:2013fsa}
with default CMS card settings (used to represent a general purpose collider detector).  
The inner tracking region is taken to extend to 130 cm from the beam line, 
the ECAL is between 130 cm and 150 cm, and the HCAL begins at 150 cm.
Jets are reconstructed with \textsc{FastJet~3.0.1}~\cite{Cacciari:2011ma}
 using the anti-$k_T$ clustering algorithm~\cite{Cacciari:2008gp} with 
jet size parameter $R=0.4$.  
Only the leading reconstructed jet with $p_T$ of at least 20 GeV 
that is 
within $\Delta R \equiv  \sqrt{(\Delta \eta)^2 + (\Delta \phi)^2} < 0.4$ of one of the
\textsc{MadGraph5} hard scattering final state quarks is used from each event 
in the analyses presented below. 
These requirements reject events with excessive radiation from final state 
quarks produced in the hard scattering process, as well as any events with 
high jet multiplicity and anomalously 
low $p_T$ per jet. 

The default \textsc{Delphes~3.4.1} detector simulator makes a number of 
simplifying assumptions to treat long-lived metastable hadrons produced in collision events. 
This includes treating all particles that originate inside the inner tracking region
 effectively as prompt coming from the interaction point. 
 In addition, 
 while 
 metastable hadrons that decay outside of 
 the inner tracking region deposit all of their energy 
 into the 
 calorimeter in \textsc{Delphes~3.4.1}, 
 whether or not these decays take place before or after the ECAL is not taken into account. 
 Finally, \textsc{Delphes~3.4.1}  ignores the decays-in-flight of metastable short-lived neutral kaons, $K_S$, 
 and strange baryons, and instead treats them as stable particles that deposit their 
 energies in both the ECAL and HCAL with a fixed relative ratio. 
 These simplifying assumptions unfortunately mask some crucial 
 observable differences between 
strange- and down-quark jets. 
So for the studies undertaken in this paper, the treatment of how long-lived metastable hadrons 
and their decay products 
are handled within the \textsc{Delphes} detector simulator is modified and improved in a number of important ways.

\begin{itemize}
 
\item {\it Reconstruction of charged particles.} 
Detector stable charged particles (with average decay lengths much larger than relevant detector dimensions) 
are taken to be the charged pion and kaon, electron, muon, and proton, 
$\pi^+ \! , K^+ \! , \, e, \, \mu, \, p$ 
(and their anti-particles). 
Other charged particles are considered metastable. 
All detector stable charged particles that originate from the prompt interaction point, or from decays-in-flight that occur out to a transverse distance of 50 cm from the beam axis, are included 
in the \textsc{Delphes} reconstructed charged track list. 
Charged particles that originate at distances greater than 50 cm from the beam axis are not 
included in the reconstructed track list.
This simple ansatz is meant to roughly model the relatively high efficiency with which charged particle tracks that originate at or not too far from the beam axis
and pass through many tracking layers of 
the inner tracking region can be reconstructed. 
The exclusion of tracks that originate from inflight decays beyond 50 cm from the beam axis 
is a rough representation of the   
rapidly falling efficiency for reconstructing charged tracks that pass through a smaller number of tracking layers
(that extend out to 130 cm). 

\item {\it Electromagnetically showering particles and the calorimeter.} 
Electromagnetically showering particles are taken to be the photon and electron, $\gamma, e$ 
(and the positron anti-particle). 
All electromagnetically showering particles that originate from the prompt interaction point, or from 
decays-in-flight up to a distance of 150 cm from the beam axis, deposit all their energy in the ECAL. 
Electromagnetically showering particles that originate at distances greater than 150 cm from the beam 
axis deposit no energy in the ECAL. 
This simple ansatz is meant to roughly model the high efficiency with which electromagnetically showering 
particles that originate either before or in the ECAL region 
deposit energy in the ECAL (which extends from 130 cm to 150 cm). 
It neglects a transition region for such particles that originate from decays-in-flight
near the outer edge of the ECAL for which only a fraction of the energy is deposited in the ECAL 
with the remaining energy punching through to the HCAL. 
However, this transition region is on average small if the ECAL is many radiation lengths in depth 
(which is the case for a general purpose collider detector). 

\item {\it Hadrons and the calorimeter.} 
Detector stable hadrons 
(with average decay lengths much larger than relevant detector dimensions) 
are taken to be the charged pion and kaon, neutron, and proton, 
$\pi^+ \! , K^+ \! , \,  n, \, p$
 (and their anti-particles)
and the long-lived neutral kaon, $K_L$. 
Other hadrons are considered metastable. 
All detector stable hadrons that originate from the prompt interaction point, or from decays-in-flight that occur out to a transverse distance of 150 cm from the beam axis, 
deposit all their energy in the HCAL (that begins at 150 cm).  
In addition, all metastable hadrons that do not decay within 150 cm of the beam axis 
also deposit all their energy in the HCAL. 
This simple ansatz is meant to model the high efficiency with which hadrons
that reach the HCAL deposit energy there, and the relatively small efficiency to 
deposit energy in the ECAL.   
Hadronic energy deposited in the ECAL is sub-dominant to that deposited in the HCAL 
if the ECAL is a fraction of interaction length in depth (which is generally the case for a 
general purpose detector).

\item {\it Treatment of metastable $K_S$ decays.} 
The final addition to \textsc{Delphes} is a new variable that corresponds to reconstruction 
of the momenta of individual short-lived neutral kaons that decay in-flight to a pair of charged pions, 
$K_S \to \pi^+ \pi^-$. 
The ability to reconstruct this decay-in-flight as a displaced, isolated 
secondary vertex with an emerging positively and negatively charged track
pair with invariant mass consistent with that of the short-lived neutral kaon 
has been well established by both the 
CMS \cite{Khachatryan:2010pw} and ATLAS \cite{Aad:2011hd} collaborations. 
This decay-in-flight is in fact now used as a standard candle to measure the efficiency of reconstructing 
displaced charged particle tracks and secondary vertices.  
Although reconstruction of individual $K_S$ momenta from   $K_S \to \pi^+ \pi^-$ decays-in-flight
within jets has not yet (to our knowledge) been explicitly utilized in any light-flavor jet identification 
algorithm, it could be included as a jet observable.  
In order to accommodate this interesting possibility, the momenta of short-lived kaons
that decay in-flight by $K_S \to \pi^+ \pi^-$ with a lab-frame decay length from the beam 
axis between 0.5 cm and 50 cm are included as additional  
observables 
with a very conservative
40\% efficiency. 
This simple ansatz is meant to roughly model the ability to reconstruct 
these decays (within jets)
from measurement of displaced charged track pair momenta 
in the inner tracking region with moderate efficiency.  
The requirement of a displaced charged track pair vertex consistent with the kaon mass 
at least 0.5 cm from the beam axis avoids confusion with prompt tracks that originate 
from the interaction point, and the exclusion of vertices beyond 50 cm from the beam axis 
is a rough representation of the   
rapidly falling efficiency for reconstructing charged tracks that 
pass through a smaller number of tracking layers
(that extend out to 130 cm). 

\end{itemize}

We note that the default \textsc{Pythia8} and \textsc{Delphes} treatment of neutral kaons (left unmodified in this work) is to propagate $K_L$ and $K_S$ as mass eigenstates.
This neglects oscillation including CP violating effects which are small and unimportant  
in the observables utilized below.  
Regeneration of $K_S$ from $K_L$ within detector material (and vice versa) is also a small 
effect and is neglected. 

For the detector based studies detailed below, four types of observable variables are 
extracted from the modified \textsc{Delphes} detector simulator described above. 
The first two are the neutral electromagnetic and neutral 
hadronic calorimeter tower energies, $E_N$ and $H_N$, 
obtained from the \textsc{Delphes} variables \texttt{eFlowPhotons} and 
\texttt{eFlowNeutralHadrons} respectively. 
The neutral electromagnetic tower energy, $E_N$, is equal to the total electromagnetic energy deposited in 
an ECAL tower minus the sum $p_T$ of all electrons and positrons that deposit energy in the tower.  
This corresponds effectively then to the total energy deposited by photons in an ECAL tower.  
The neutral hadronic tower energy, $H_N$, is equal to the total hadronic energy deposited 
in an HCAL tower minus the sum $p_T$ of all charged hadrons (which amounts to all charged particles 
except electrons and muons and their anti-particles) that deposit energy  in the tower. 
This corresponds effectively then to the total energy deposited by neutral hadrons in an HCAL tower. 
These calorimeter tower variables are meant to roughly model similar variables 
that form the basis of particle flow algorithms utilized in some analyses of actual LHC detector data. 

The final two variables used in the detector based studies below are charged track based. 
The first of these, $T$, is the sum $p_T$ of all detector stable 
charged particles that point at a calorimeter tower. 
Within a light-flavor jet the electron and muon content is generally small, so for the studies carried out here 
$T$ gets contributions almost solely from detector stable charged hadrons. 
In a more generally setting the contributions to $T$ from electron objects and muons could be subtracted
to obtain essentially the same variable. 
The final variable, $K_{S_{\pi^+ \pi^-}}$, is the sum $p_T$ of all short-lived kaons that point at a calorimeter 
tower and decay to a charged pion pair, $K_S \to \pi^+ \pi^-$, within the lab-frame decay distance window 
and efficiency described above.  
These charged track based variables are meant to roughly model what could be achieved 
by utilizing individual reconstructed charged tracks in the inner tracking region of an actual LHC detector. 
The charged track variables are course grained in the transverse directions to the beam axis on 
the same scale as individual calorimeter towers just to simplify the analysis.  

For jet-level observables used in the detector based studies 
below, the four tower observables, $E_N, H_N, \, T, \, K_{S_{\pi^+ \pi^-}}$, 
are each normalized to the \textsc{Delphes} reconstructed jet $p_T$. 
For the truth-level studies below, the $p_T$ of individual particles that form 
a \textsc{Delphes} reconstructed jet
are taken from the \textsc{Pythia8} particle list 
and normalized to the reconstructed jet $p_T$.


\section{Strange versus Down Jets}
 \label{sec:strangejets}
 
Strange- and down-quarks have identical strong and electromagnetic interactions, 
and both have bare masses (just) below the QCD scale.  
So in the perturbative QCD regime, the 
showering of strange- and down-quarks produced in hard scattering or from heavy decay processes 
are essentially identical. 
Quark flavor is preserved during showering and hadronization, so the main difference 
is that the hadrons within 
a jet initiated by a hard strange-quark must carry a net strangeness of one unit, 
whereas the hadrons within a jet initiated by a hard down-quark must 
have one unit of down-ness, but no net strangeness.  
The specific particle content of strange- and down-quark jets that follows from this difference, 
and the degree 
to which the difference can be observed and exploited in the current generation 
of general purpose collider detectors are 
detailed in the sub-sections below.


\subsection{Constituents of Strange and Down Jets} 

The hadrons that materialize from the showering and hadronization process within 
a high momentum light-flavor jet are predominantly charged and neutral pions, 
$\pi^\pm  , \pi^0$, because of the much lighter masses compared with other hadrons. 
The charged pion decay length is much larger than a collider detector, and so is a detector stable particle, 
whereas the neutral pion decays promptly to two photons, $\pi^0 \to \gamma \gamma$. 
So the predominant observable constituents of light-flavor jets in a collider detector 
are detector stable charged pions and photons, both of which originate from the primary 
interaction point. 

The net unit of strangeness within a strange-quark jet, compared with the net unit of 
down-ness within a down-quark jet, is carried by definition in each case, by a single hadron 
among the more numerous charged and neutral pions and other hadrons. 
Even so, this can 
 lead on average 
to important qualitative differences between these types of jets. 

The single net unit of down-ness within a jet initiated by a high momentum down-quark 
after showering, hadronization, and any hadronic decays, can reside with greatest probability  
in a charged pion, $\pi^-$, or with a lower probability because of larger masses, 
in a proton or neutron baryon, $p,n$, 
all of which are detector stable particles. 

In contrast, 
the single net unit of strangeness within a jet initiated by a high momentum strange-quark 
after showering, hadronization, and any prompt decays of highly unstable hadrons, 
can reside with greatest probability in either a kaon linear combination of long- and short-lived 
neutral kaon mesons, 
$K^0 \simeq (K_L - K_S)/\sqrt{2}$, 
or 
a charged kaon meson, $K^-$, or with lower probability because of larger masses,  
a neutral or charged strange baryon within the baryon octet, 
$\Lambda^0, \Sigma^+, \Sigma^0, \Sigma^-,\Xi^0, \Xi^- $.
All of these meson and baryon strangeness carrying states, 
except $\Sigma^0$,  are relatively long lived, with strangeness violating 
decays only through weak flavor-changing interactions. 
The charged kaon, $K^-$, and the 
long-lived component of the neutral kaon, $K_L$, 
have decay lengths of 370 cm and 1500 cm respectively, and so are effectively detector stable particles
in a high energy collider detector. 
The short-lived component of the neutral kaon, $K_S$, and the metastable strangeness one baryons 
$\Lambda^0, \Sigma^+, \Sigma^-$, have decay lengths of 
2.7 cm and 7.8 cm, 2.4 cm, 4.4 cm respectively, and can decay in-flight 
within the detector volume, 
while the remaining strangeness one baryon, $\Sigma^0$, undergoes prompt radiative decay to $\Lambda^0$. 
The strangeness two baryons $\Xi^0, \Xi^-$ have decay lengths of 8.7 cm, 4.9 cm, 
and also can decay in-flight within the detector volume. 
Of these short-lived metastable strangeness carrying states, the short-lived kaon decays to a pair 
of charged or neutral pions, $K_S \to \pi^+ \pi^-, \pi^0 \pi^0$, with branching 
ratios of 69\% and 31\% respectively. 
The metastable strangeness one and two baryons (cascade) decay to proton or neutron baryons and charged 
or neutral pions.


\begin{figure}[htbp]
  \begin{minipage}{0.48\hsize}
  \begin{center}
   \includegraphics[clip, width=8.5cm]{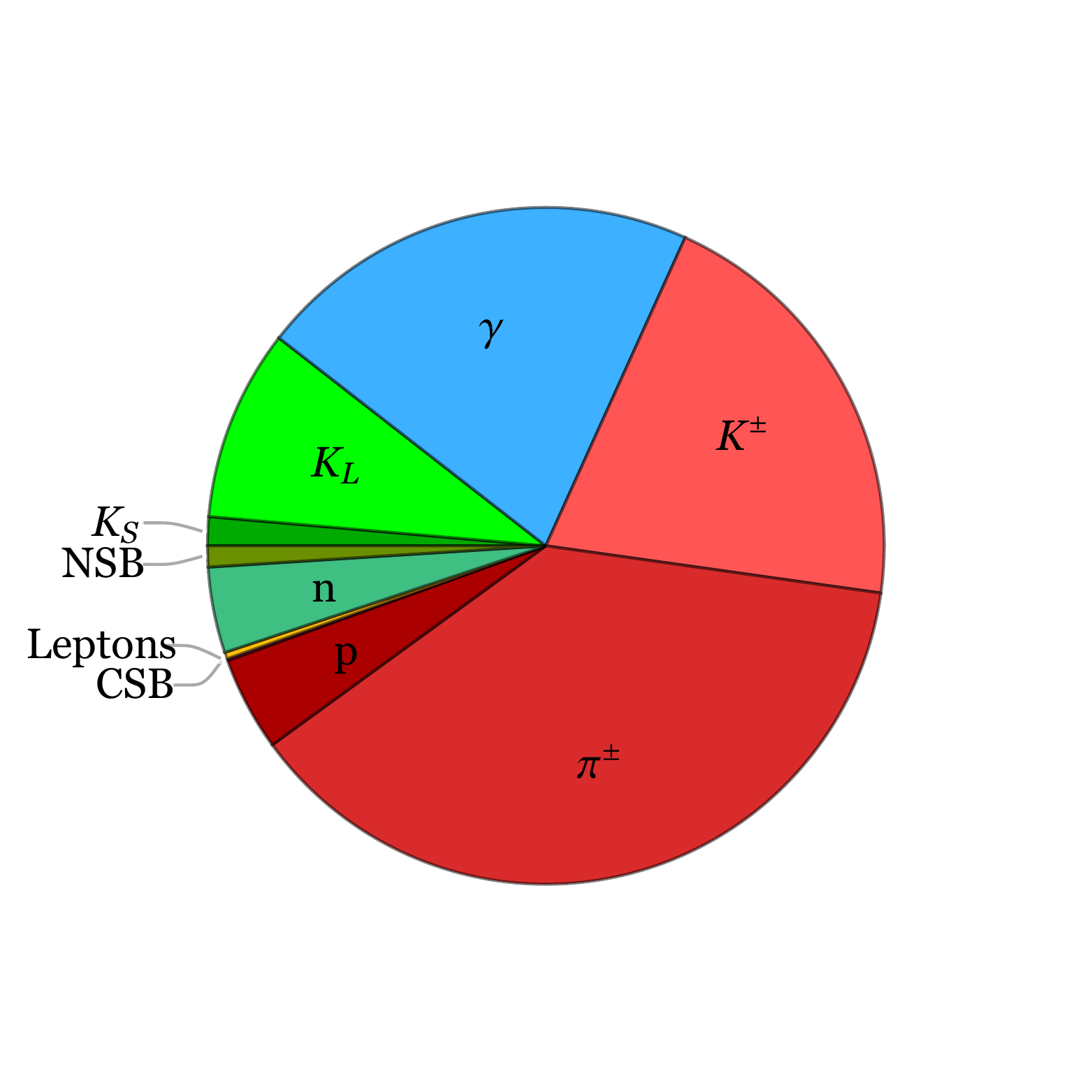}
  \end{center}
  \vspace{-0.45in}
  {$~~~~~~~~~~~~~~~~{\rm Strange}~~ p_T = 45 \, \rm GeV$}
  \end{minipage}
 \hspace{0.5cm}
 \begin{minipage}{0.48\hsize}
  \begin{center}
   \includegraphics[clip, width=8.5cm]{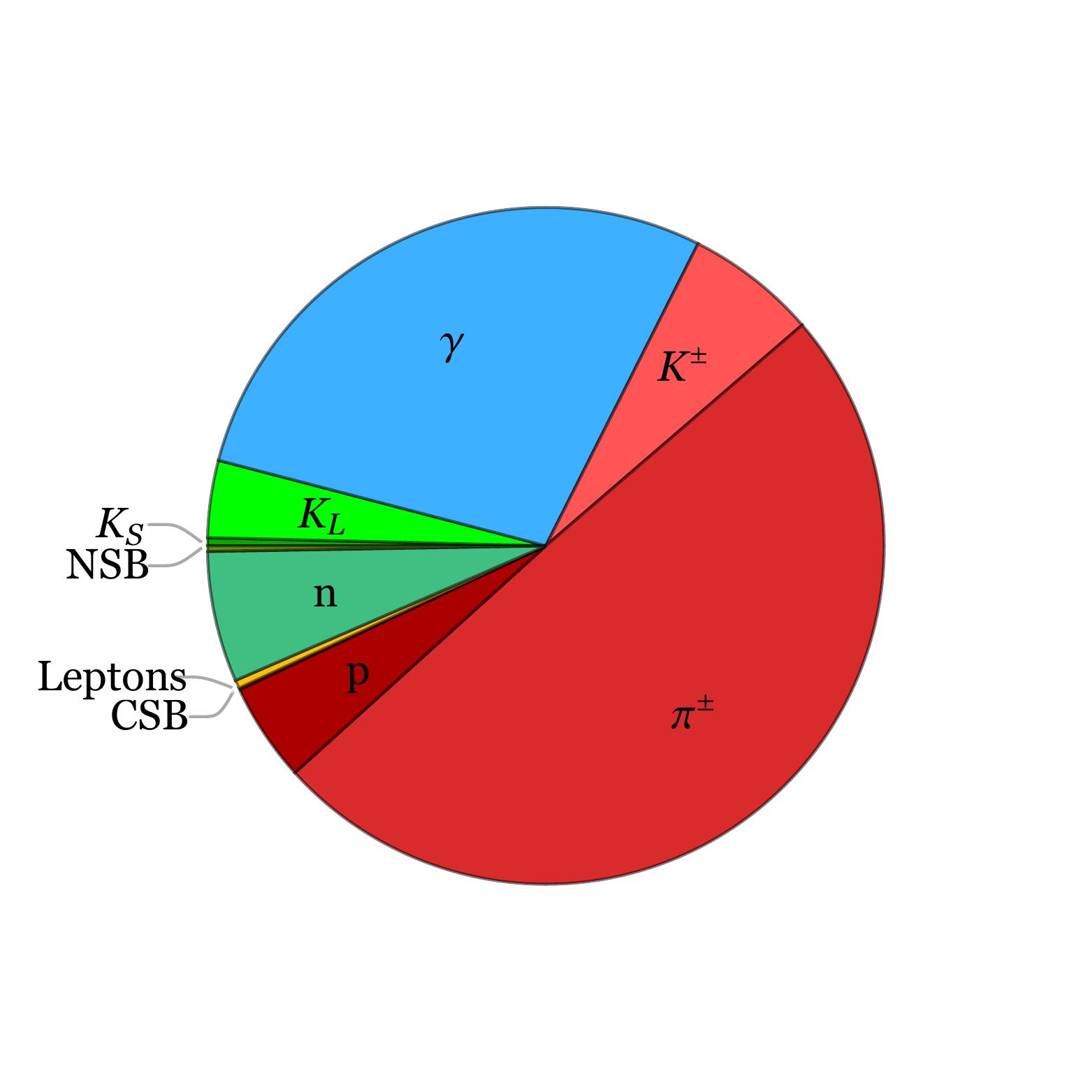}
  \end{center}
  \vspace{-0.45in}
 {$~~~~~~~~~~~~~~~~{\rm Down}~~ p_T = 45 \, \rm GeV$} 
 \end{minipage}
 \vspace{0.30cm}
   \begin{minipage}{0.48\hsize}
  \begin{center}
   \includegraphics[clip, width=8.5cm]{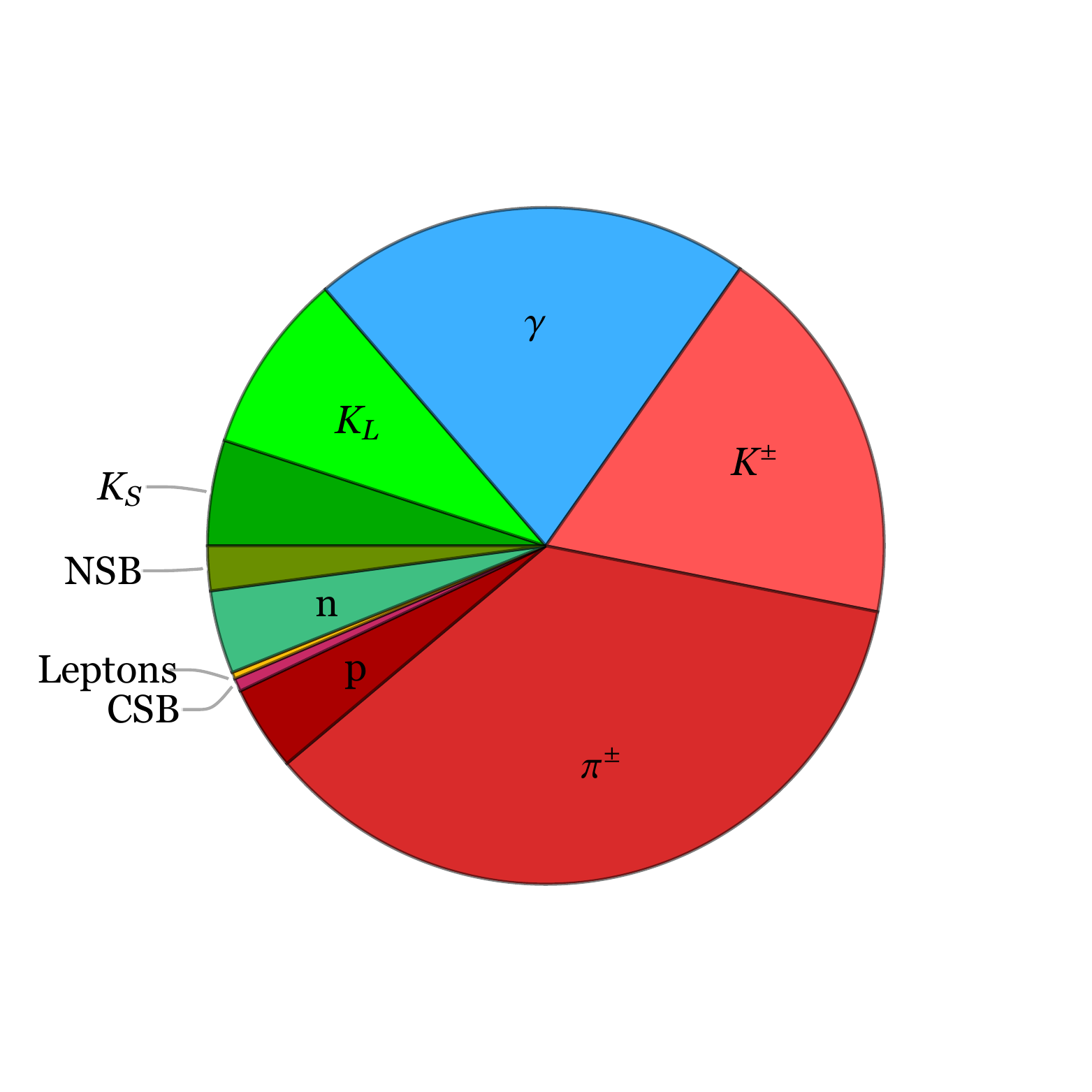}
  \end{center}
  \vspace{-0.45in}
 {$~~~~~~~~~~~~~~~~{\rm Strange}~~ p_T > 200 \, \rm GeV$}
  \end{minipage}
 \hspace{0.5cm}
 \begin{minipage}{0.48\hsize}
  \begin{center}
   \includegraphics[clip, width=8.5cm]{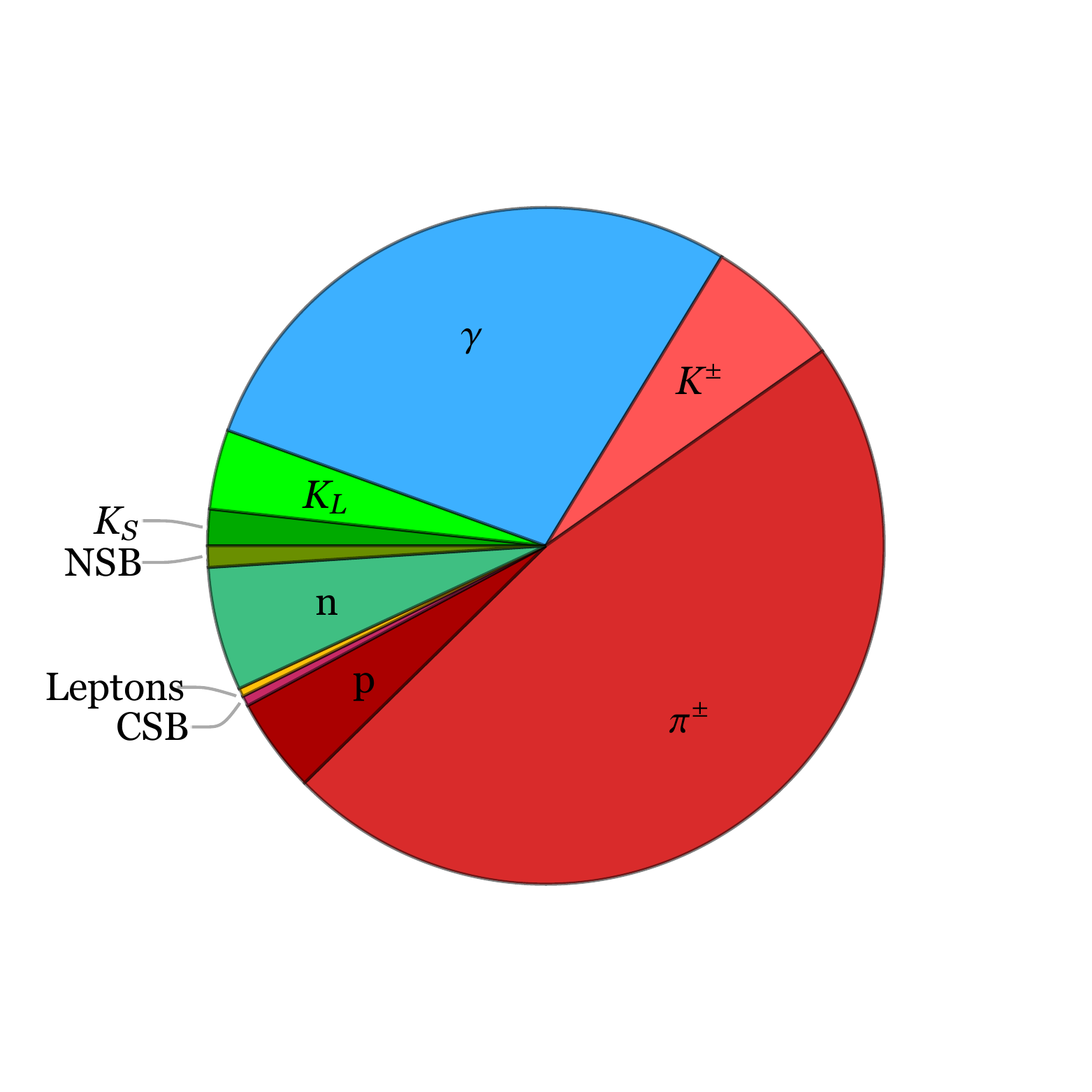}
  \end{center}
   \vspace{-0.45in}
 {$~~~~~~~~~~~~~~~~{\rm Down}~~ p_T > 200 \, \rm GeV$}
 \end{minipage}
 \vspace{0.5cm}
  \caption{
  Total momentum fractions carried by effectively detector stable particles 
  that reach the calorimeters
in strange- or down-quark jets arising from either $Z$-boson decay
with central jet $p_T$ = 45 GeV, or from QCD initiated 
13 TeV proton-proton central collisions with jet $p_T > 200$ GeV, 
simulated with \textsc{Pythia8}. 
Here effectively detector stable is  
either a detector stable particle
 or a metastable particle that does not decay within 
 a transverse distance of 150 cm from the beam line. 
Red and green correspond to electrically charged and neutral particles, and blue to photons. 
NSB and CSB refer to neutral and charged strange-baryons. 
  }
  \label{fig:jetclassification}
\end{figure}

An important consideration in the practical utility of exploiting the differences between 
strange- and down-quark jets outlined above is the momentum fraction carried by 
the various types of detector stable and meta-stable particles within the jets.  
In order to visually illustrate these differences the truth-level 
average $p_T$ weighted momentum fractions 
obtained from \textsc{Pythia8}
for each type of effectively detector 
stable particle within the leading jet 
in the four sets of event samples
described in the previous section are shown in the pie charts in Fig.~1. 
In this figure, effectively detector stable is defined to be a either 
detector stable particle or a meta-stable particle that does not decay within 
a transverse distance of 150 cm from the beam axis. 
Such particles reach and deposit energy in the calorimeters of the canonical
general purpose collider detector considered here.  
All effectively detector stable 
particles within a jet that satisfy these criteria contribute to the momentum weighted fractions 
in the pie charts, including ones that originate directly from hadronization at the 
primary interaction vertex, as well as ones that emerge from secondary vertices
from 
decays-in-flight of metastable particles.  
Red and green in the pie charts correspond to electrically charged and neutral particles 
respectively, and blue refers to photons. 

A number of patterns evident in the pie charts in Fig.~1 are worth highlighting. 
In all cases for either strange- or down-quark jets, the momentum weighted fraction of 
charged kaons is twice that of long-lived neutral kaons.  
The is because the hadronization probability for charged and neutral kaons is essentially identical, 
and long-lived neutral kaons make up half of the neutral kaon content at hadronization. 
The charged kaons and long-lived neutral kaons are both detector stable particles and 
appear directly in the pie charts.  
In contrast, only a fraction of the short-lived neutral kaons that make up the other half of the neutral kaon content
from hadronization 
reach the calorimeters at a transverse distance of 150 cm from the beam axis and appear in the pie charts, 
with the remainder undergoing decays-in-flight. 
The fraction of short-lived kaons that reach the calorimeters 
is directly proportional to the average lab-frame decay length, which 
in turn is proportional to the total transverse momentum of the jet. 
For $p_T = 45$ GeV jets, the short-lived kaon momentum weighted fraction that reaches the calorimeters 
is rather small, while for $p_T > 200$ GeV jets, it is a sizable fraction of that of the long-lived kaons.  
The momentum weighted fraction of neutral strange baryons that reaches the calorimeters for 
$p_T > 200$ GeV jets is also larger than that for $p_T = 45$ GeV jets for the same reason. 

The charged kaon momentum weighted fractions in strange-quark jets 
shown in the pie charts on the left-hand side of Fig.~1
are more than three times larger 
than that in down-quark jets shown on the right-hand side.  
However, the sum of the charged kaon and charged pion momentum weighted fractions 
at hadronization are essentially identical. 
This is because the hadronization probability into charged particles for the net unit of strange-ness or down-ness 
within a strange- or down-quark jet respectively 
is roughly equal and 
dominated by these lightest net strange-ness and down-ness carrying mesons. 
The near equality 
of the sum of the charged kaon and charged pion momentum weighted fractions 
is evident in comparing the pie charts for strange- and down-quark jets with $p_T > 200$ GeV in Fig.~1. 
For $p_T = 45$ GeV 
the sum  
is slightly larger for strange-quark jets than for down-quark jets, with the excess 
coming mainly from decays-in-flight of short-lived neutral kaons to charged pion pairs. 
The current generation of general purpose collider detectors 
can not effectively distinguish charged kaons from charged pions. 
So the near equality of the sum of the momentum weighted fractions of these particles 
means that the individual 
differences can not be exploited as a handle to distinguish strange- and down-quark jets 
in these detectors. 

The momentum weighted fraction of long-lived neutral kaons 
in strange-quark jets is more than two times larger than in down-quark jets. 
This is because a long-lived neutral kaon can carry the net unit of strange-ness in 
a strange-quark jet, but can not carry the net unit of down-ness in a down-quark jet. 
The difference is apparent in comparing the pie charts for strange- and down-quark jets with 
$p_T = 45$ GeV as well as for $p_T > 200$ GeV in Fig.~1. 
In a collider detector 
long-lived neutral kaons deposit energy primarily in the HCAL.  
So the excess fractional energy deposited there by the net unit of strangeness 
in a strange-quark jet provides an observable handle with which on average to distinguish 
these from down-quark jets. 
However, the  momentum weighted neutron fraction is somewhat larger in
down-quark jets than in strange-quark jets. 
This is because the net unit of down-ness in a down-quark jet can, with some small probability, 
be carried by a neutron. 
Since neutrons also deposit energy primarily in HCAL, this tends on average 
to reduce the effectiveness of neutral hadronic energy alone to distinguish 
strange-quark jets from down-quark jets. 

The final important feature 
evident in the pie charts in Fig.~1 is that 
the photon momentum weighted fractions 
in strange-quark jets 
shown in the pie charts on the left are smaller than those in the down-quark jets shown on the right-hand side. 
Photons within both types of jets arise predominantly from prompt decays of neutral pions to photon pairs, 
and to a lesser extent from prompt eta and eta prime decays to photons, either directly or through
cascade decays with intermediate neutral pions, 
and decays-in-flight of short-lived kaons to photons through intermediate neutral pion pairs. 
None of these states except the short-lived kaon carry net strangeness or down-ness. 
However, it is possible for the primary hard strange-quark 
to hadronize 
into one of these non-strangeness carrying states, with the net strange-ness or down-ness carried by 
a secondary state of lower momentum fraction. 
In a down-quark jet the probability that the 
primary down-quark from the hard process hadronizes into a charged pion is twice that 
into a neutral pion. 
In contrast, in a strange quark jet the only states that are available for the primary 
strange quark to hadronize into in conjunction with an up- or down-anti-quark 
are the strangeness carrying charged and neutral kaon mesons.  
The primary strange quark hardronizes into an eta meson only in conjunction with a 
heavier anti-strange-quark. 
So the primary hard quark  
in a strange-quark jet has lower probability to 
hadronize into a state that promptly decays to photons than in a down-quark jet. 
In strange-quark jets photons also 
arise from decays-in-flight of short-lived kaons, but the total momentum weighted 
photon fraction is still smaller than that of down-quark jets.

Although not of primary importance in distinguishing strange- and down-quark jets, 
it is also worth noting that 
the sum of the strange and non-strange 
baryon momentum weighted fraction in strange-quark jets is somewhat less than 
that in down-quark jets. 
This is because the larger mass of the strange-quark results in a smaller probability for strange-quarks 
to hadronize into strange baryons relative to mesons, compared with the probability for 
down-quarks to hadronize into non-strange baryons relative to mesons.

In summary, for effectively detector stable particles, 
the momentum weighted fraction of long-lived neutral kaons are higher 
in a strange-quark jet than in a down-quark jet. 
Conversely, the momentum weighted fraction of photons is smaller in 
a strange-quark jet than in a down-quark jet. 
And the total momentum weighted fraction in detector stable charged particles is 
very similar in both types of jets.


\subsection{Strange and Down Jet Signatures in a Collider Detector}


\begin{figure}[t]
  \begin{center}
   \includegraphics[clip, width=9cm]{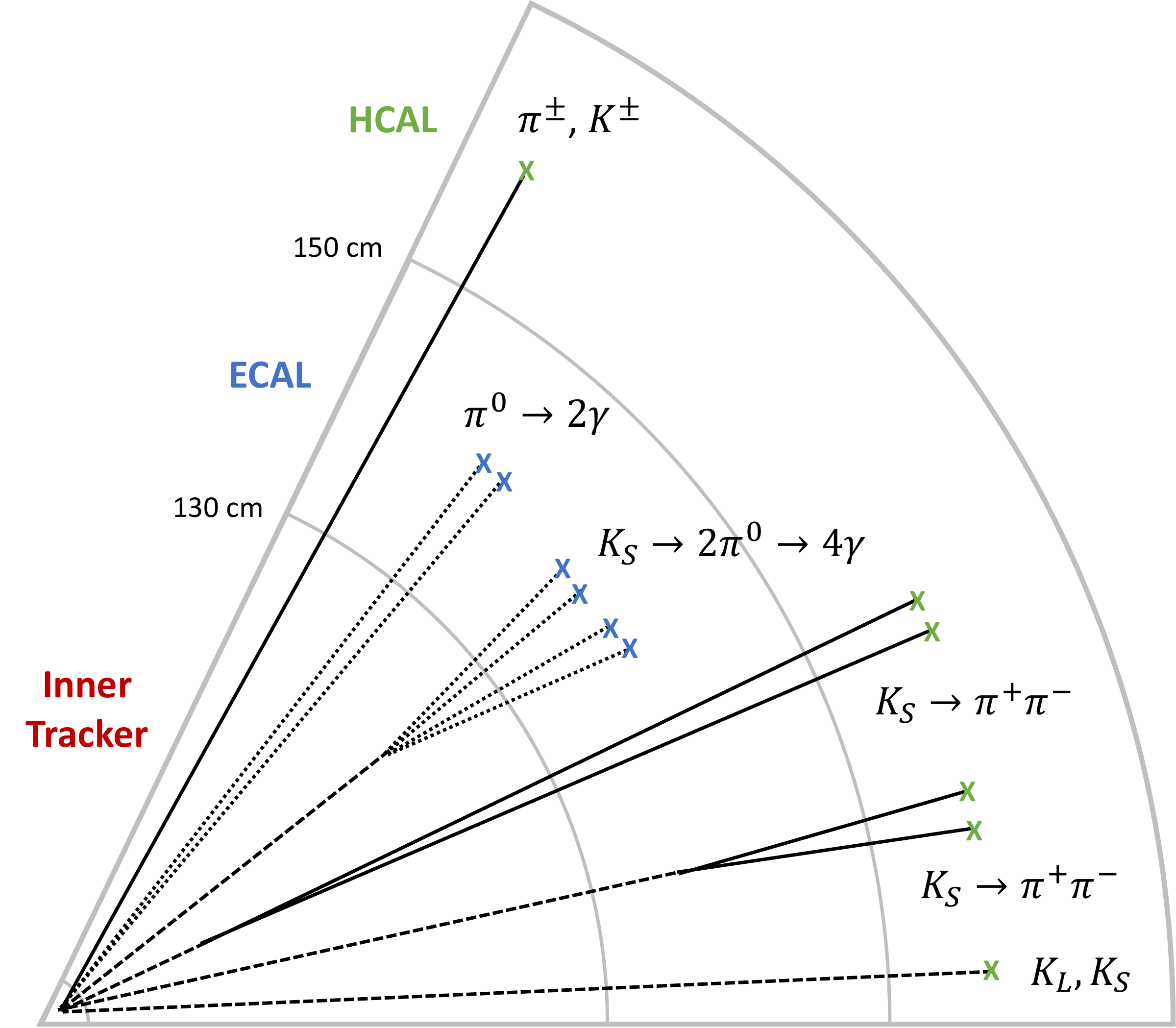}
   \vspace{0.5cm}
  \end{center}
  \caption{ Cartoon (not to scale) of pions and kaons from central collisions 
  traversing a transverse wedge of a 
  general purpose high energy collider detector. 
  Solid lines are charged particles, dashed lines are neutral particles, 
  and dotted lines are photons.  
  An $\times$ indicates that the particle is absorbed or stops 
    within the calorimeter subsystem.  
    Neutral pions decay promptly to photons that are absorbed in the electromagnetic 
    calorimeter.  
  Charged pions and kaons and long-lived neutral kaons are detector stable particles
  that are stopped in the hadronic calorimeter. 
  Short-lived neutral kaons can decay in flight to neutral or charged pion pairs 
  within any of the detector subsystems shown,
  or can be stopped before decay in the hadronic calorimeter.   
}
  \label{fig:Detector_Kshort}
\end{figure}

The response of a general purpose collider detector to pions and kaons originating 
from the primary interaction point of a high energy collision
are illustrated in Fig~2. 
The charged pion and kaon, $\pi^-, K^-$, are both detector stable particles. 
The momenta of these particles are measured in the inner tracking region, 
and their energy is deposited primarily in the HCAL. 
These particles are essentially indistinguishable in the current generation of general purpose 
collider detectors.   
Neutral pions, $\pi^0$, decay promptly to photon pairs, $\pi^0 \to \gamma \gamma$, 
that are absorbed in the ECAL. 
Long-lived neutral kaons are detector stable and deposit energy primarily in the HCAL. 
Short-lived neutral kaons can either 
reach the HCAL and deposit energy there before decaying, 
or undergo decays-in-flight to neutral or charged pion pairs. 
If such decays to neutral pions pairs, $K_S \to \pi^0 \pi^0 \to \gamma \gamma \gamma \gamma$, 
take place within the inner tracking region or ECAL, the resulting photons are absorbed 
in the ECAL. 
Decays of short-lived kaons to charged pion pairs 
within the ECAL, $K_S \to \pi^+ \pi^-$, results in energy deposition mainly in the HCAL.  
For short-lived kaons that decay in-flight early enough within the inner tracking region to charged pions, 
$K_S \to \pi^+ \pi^-$, the momenta can be reconstructed from the charged pions tracks  
which also deposit energy primarily in the HCAL.
 
The differences of the momentum weighted fractions of effectively detector stable particles 
in strange- and down-quark jets detailed in the previous sub-section, 
along with the detector responses to the most important of these particles outlined above, 
suggest three main (correlated) observables 
which can provide handles to distinguish between these types of jets.  
The first is the jet neutral hadronic energy fraction, $H_N$, defined in section 2. 
Since strange-quark jets have 
a higher momentum fraction of long-lived neutral kaons
and short-lived kaons that reach the HCAL, 
this observable will on average be larger than for down-quark jets.  
The second is the jet electromagnetic neutral energy fraction, $E_N$.
Since strange-quark jets have a lower momentum fraction of photons, 
this will on average be smaller than for down-quark jets. 
The third observable is the short-lived kaon $K_S \to \pi^+ \pi^-$ 
decay-in-flight momentum fraction, 
$K_{S_{\pi^+ \pi^-}}$, also defined in section 2. 
Since the neutral kaon momentum weighted fraction is larger for strange-quark jets, 
this observable will on average be larger than for down-quark jets. 
The remaining observable defined in section 2 is the 
charged track momentum fraction, $T$. 
As discussed in the previous sub-section, 
the charged particle momentum weighted fraction is not significantly different  
for strange- and down-quark jets. 
So this observable is not expected to add much discriminating power between these 
types of jets.

\begin{figure}[!t]
  \begin{minipage}{0.48\hsize}
  \begin{center}
   \includegraphics[clip, width=7.5cm]{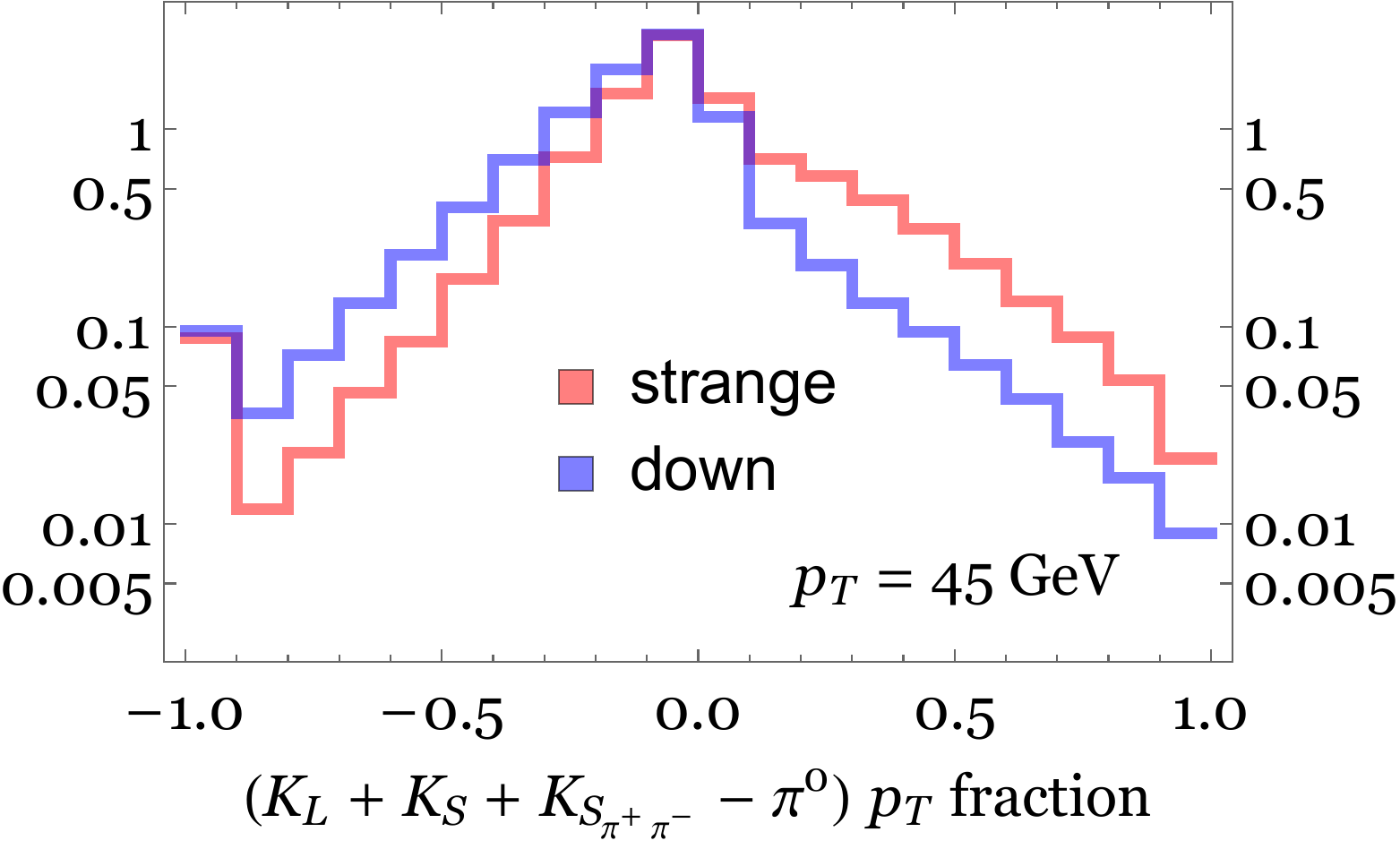}
  \end{center}
 \end{minipage}
 \hspace{0.5cm}
 \begin{minipage}{0.48\hsize}
  \begin{center}
   \includegraphics[clip, width=7.5cm]{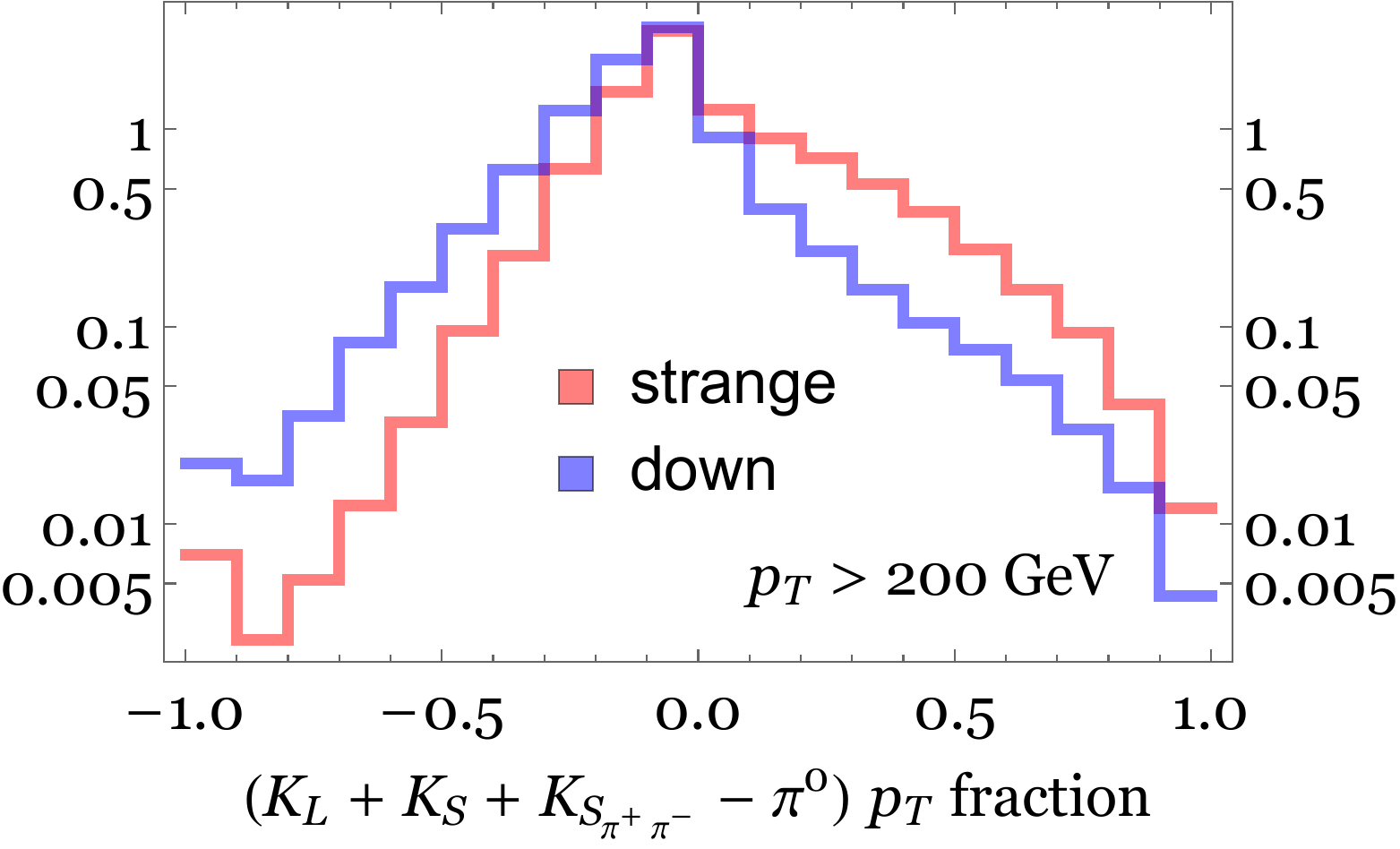}
  \end{center}
 \end{minipage}
 \vspace{0.5cm}
  \caption{
  \textsc{Pythia8} truth-level distributions (without detector simulation) of total 
    neutral kaon minus neutral pion transverse momentum fraction 
  for strange- (red) or down- (blue) quark central jets arising from either $Z$-boson decay 
  or QCD initiated 13 TeV proton-proton collisions with jet $p_T > 200$ GeV.
 Short-lived neutral kaons are only included with lab-frame decay distances greater 
   than 150 cm or with decays to charged pion pairs with a lab-frame decay distance 
   between 0.5 and 50 cm.  
 }
  \label{fig:TruthLevel}
  \end{figure}

In order to gain a measure of the ultimate utility of these observables, 
it is instructive to first consider idealized truth-level jet variables 
without full detector effects.
The change in the 
neutral kaon and photon momentum fractions 
shown in the pie charts in Fig.~1 are anti-correlated
in going from strange-quark jets to down-quark jets. 
Since the photons arise mainly from neutral pion decay, 
a simple truth-level jet variable that captures this 
correlated difference is the neutral kaon momentum fraction minus the neutral 
pion momentum fraction. 
The distributions of this truth-level variable obtained from \textsc{Pythia8}
for the leading jet in each event in the four sets of event samples
described in section 2 are shown in Fig.~3. 
In order to retain some of the most relevant irreducible detector effects, 
only short-lived neutral kaons that reach the HCAL at 150 cm from the beam axis, 
or decay to charged pion pairs  
in the inner tracking region 
with lab-frame decay distance between 0.5 and 50 cm
(and so can be reconstructed in the inner tracking region) are included in Fig.~3. 
All neutral pions and long-lived neutral kaons are included.   
Although in all cases the distributions peak at similar values of the neutral 
kaon minus neutral pion momentum fraction,  
the distributions for strange-quark jets are shifted to larger values, 
with the largest fractional differences in the tails of the distributions 
at large and small values.

\begin{figure}[!t]
  \begin{minipage}{0.48\hsize}
  \begin{center}
   \includegraphics[clip, width=7.5cm]{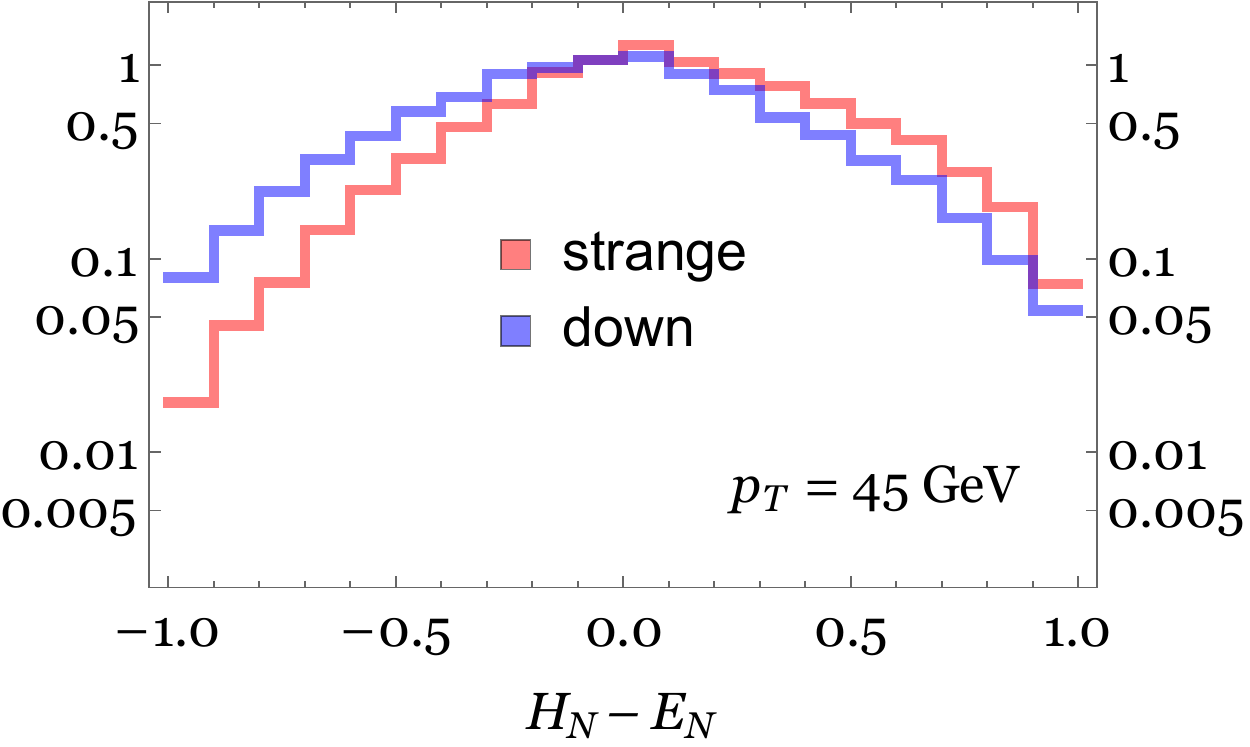}
  \end{center}
 \end{minipage}
 \hspace{0.5cm}
 \begin{minipage}{0.48\hsize}
  \begin{center}
   \includegraphics[clip, width=7.5cm]{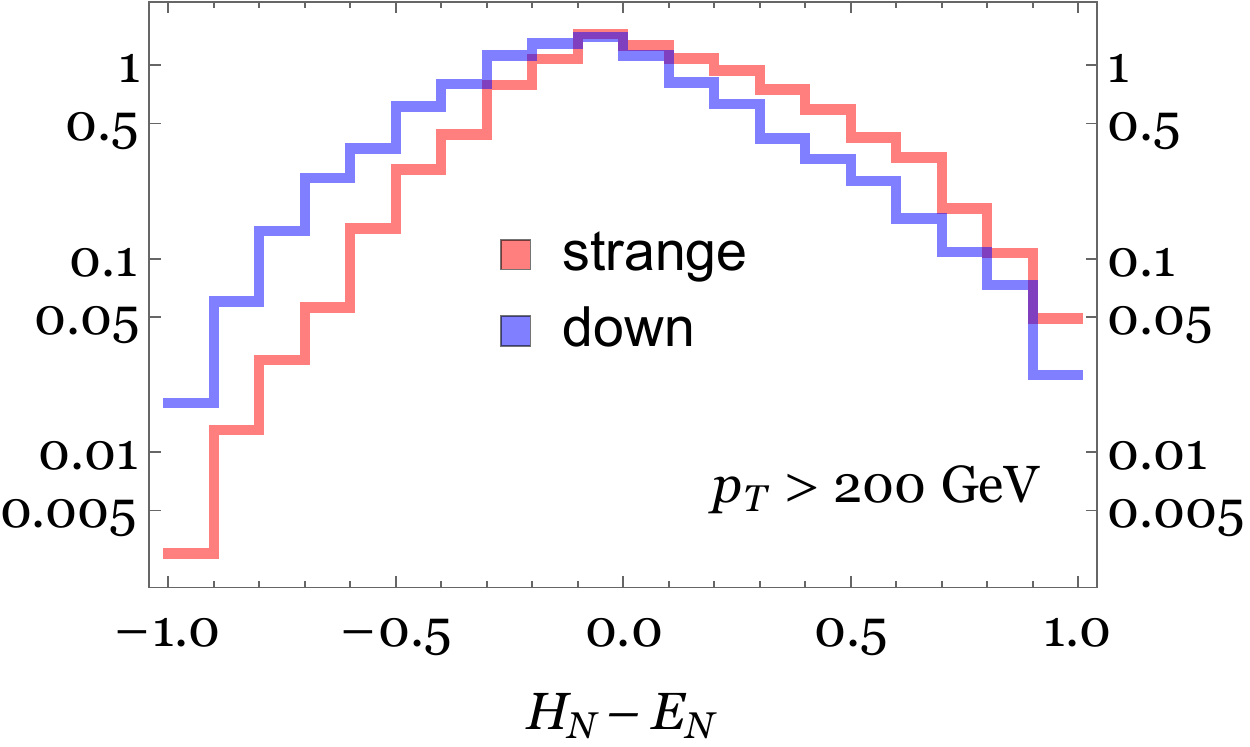}
  \end{center}
 \end{minipage}
 \vspace{0.5cm}
  \caption{
  Distribution of jet total neutral hadronic energy fraction, $H_N$,  
  minus total electromagnetic energy fraction, $E_N$, from the 
  \textsc{Delphes} detector simulation for 
  strange (red) or down (blue) quark central jets arising from either   
  $Z$-boson decay or QCD initiated 13 TeV proton-proton central 
  collisions with jet $p_T > 200$ GeV. 
  }
\label{fig:CutBased}
  \end{figure}

The detector-level observable version of the neutral kaon minus neutral pion 
momentum fraction is the neutral hadronic energy minus neutral electromagnetic energy
of a jet, $H_N - E_N$. 
The distributions of this detector-level variable obtained from \textsc{Delphes} 
for the leading jet in each event in the four sets of event samples
described in section 2 are shown in Fig.~4. 
Although somewhat washed out compared with the distributions of the truth-level variable 
given above, the same general features are still apparent, with strange-quark jet 
distributions shifted to higher values compared with down-quark jets.

\begin{figure}[!t]
  \begin{minipage}{0.48\hsize}
  \begin{center}
   \includegraphics[clip, width=7cm]{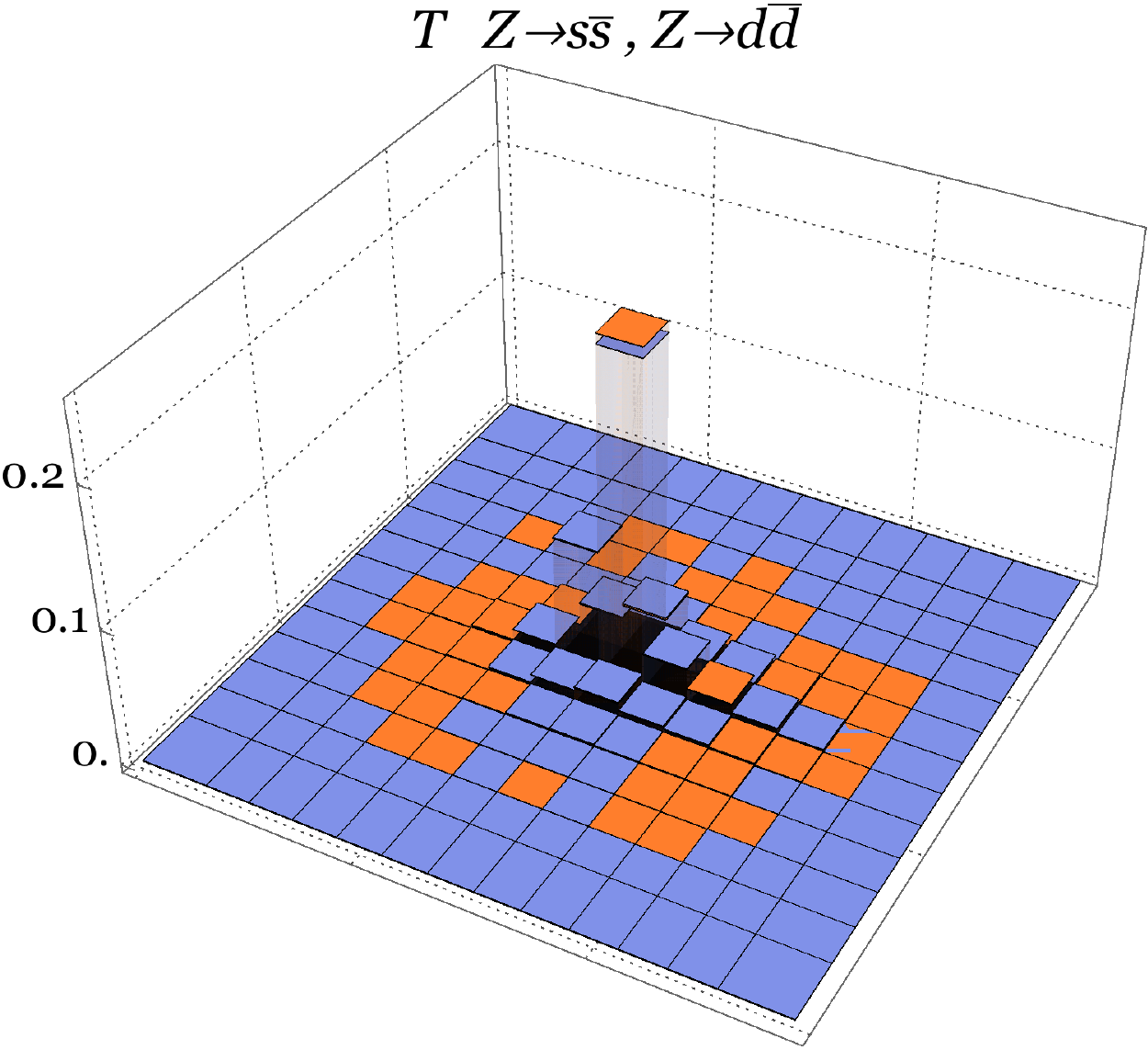}
  \end{center}
    \vspace{0.5cm}
 \end{minipage}
 \hspace{0.5cm}
 \begin{minipage}{0.48\hsize}
  \begin{center}
   \includegraphics[clip, width=7cm]{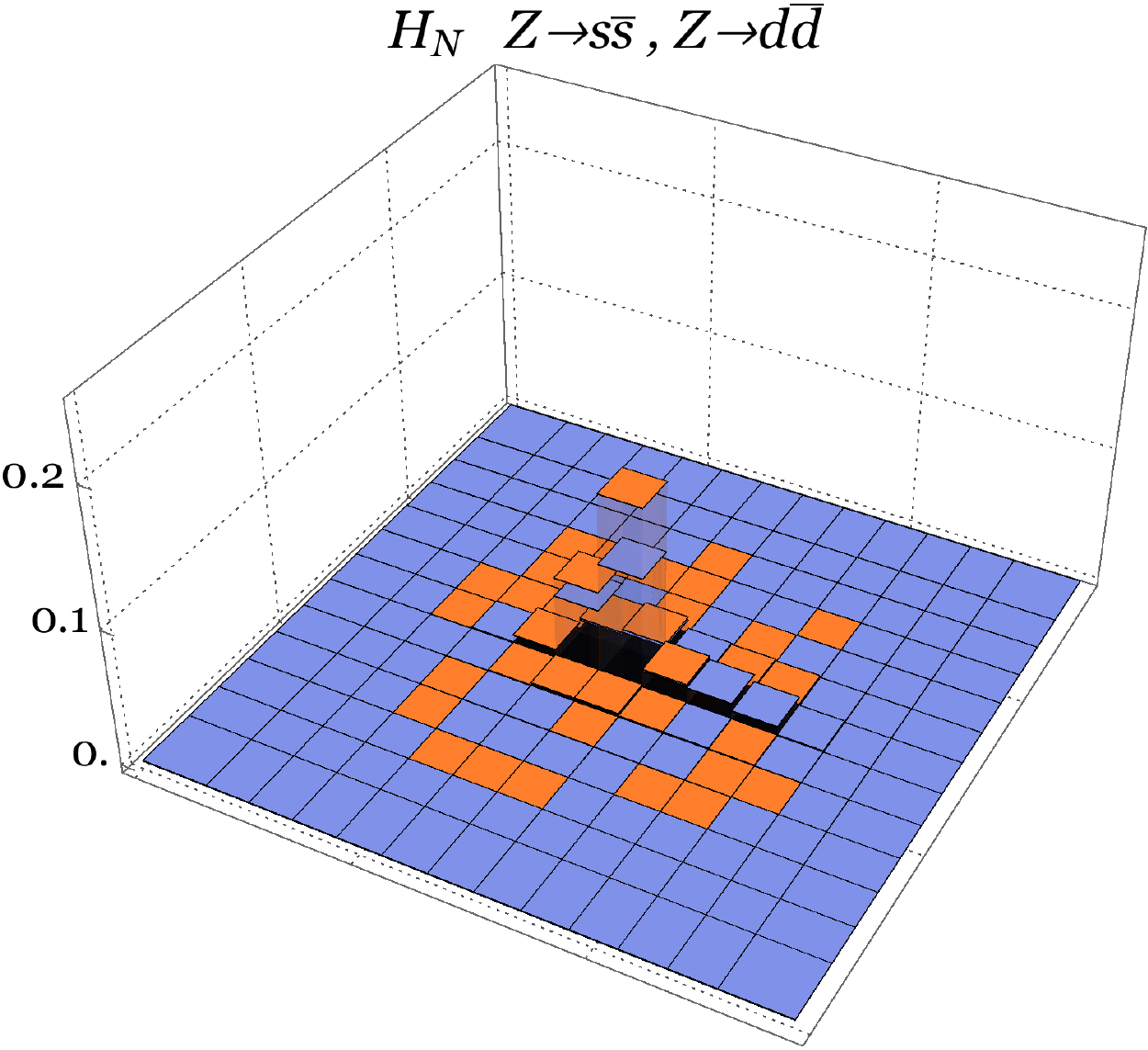}
  \end{center}
   \vspace{0.5cm}
 \end{minipage}
   \begin{minipage}{0.48\hsize}
  \begin{center}
   \includegraphics[clip, width=7cm]{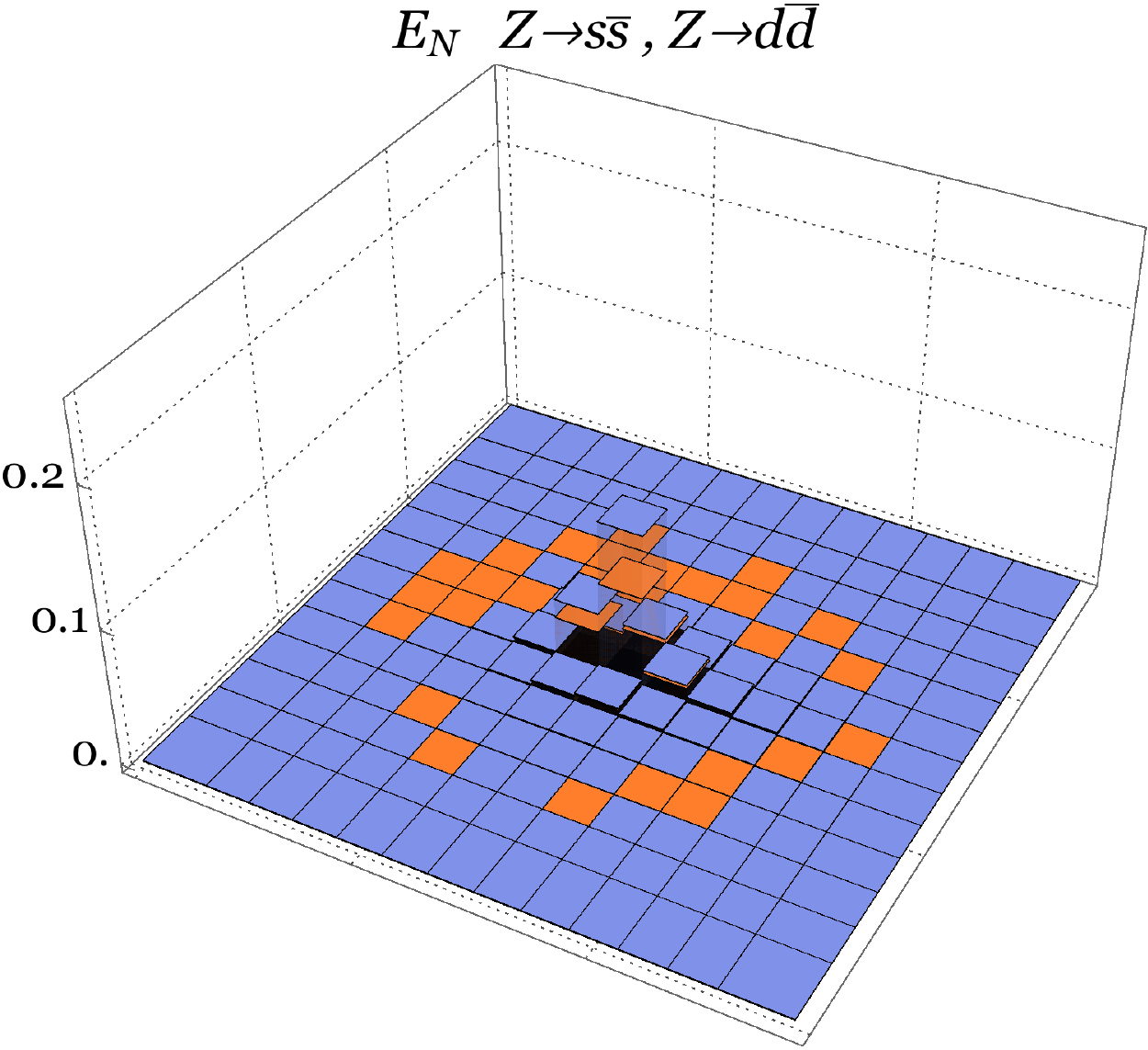}
  \end{center}
 \end{minipage}
 \hspace{0.5cm}
 \begin{minipage}{0.48\hsize}
  \begin{center}
   \includegraphics[clip, width=7.25cm]{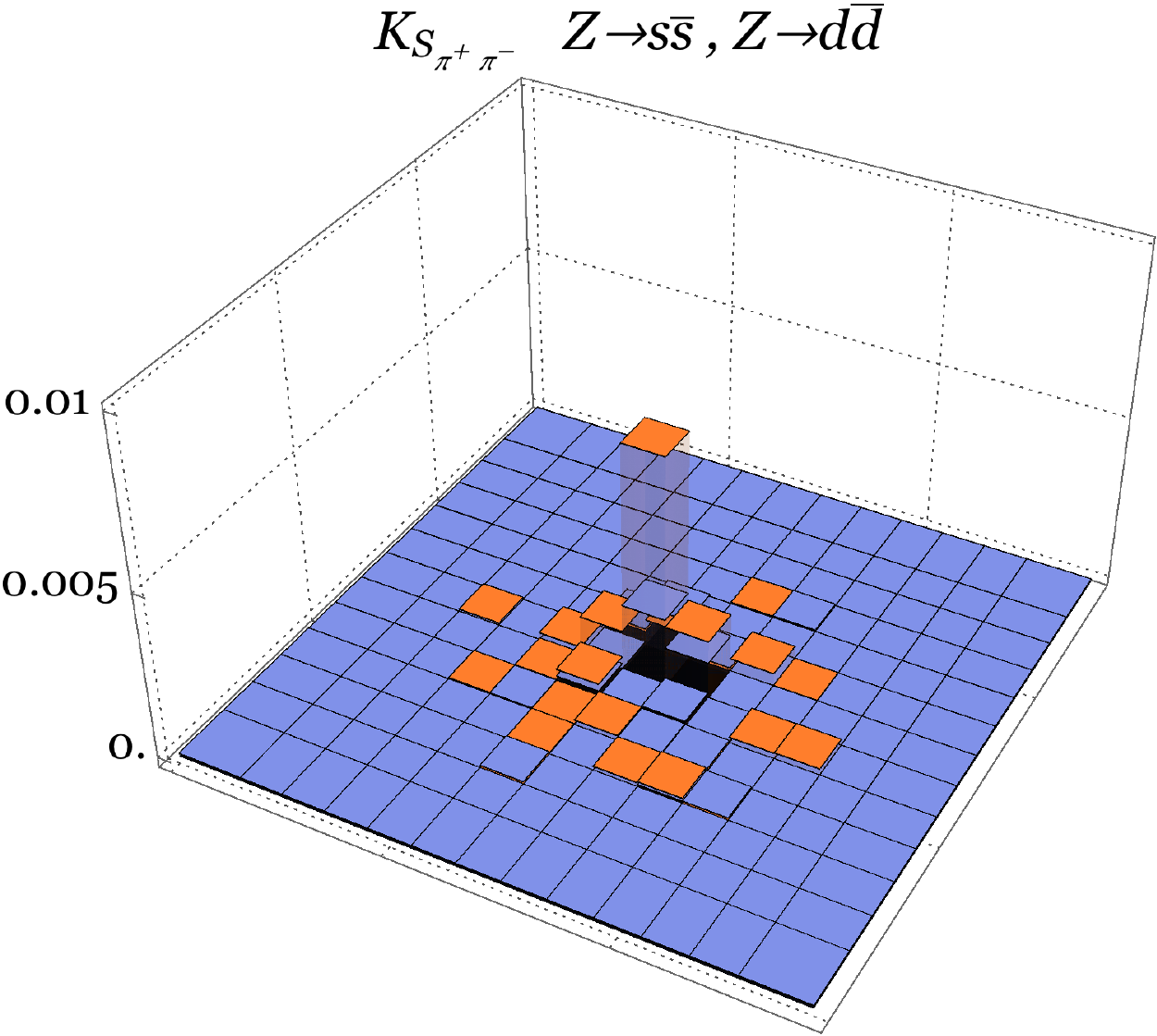}
  \end{center}
 \end{minipage}
 \vspace{0.5cm}
  \caption{Average jet
  lego plot images from 
  \textsc{Delphes} detector simulation 
    of charged track transverse momentum fraction, $T$, 
   hadronic neutral energy fraction, $H_N$, 
   electromagnetic energy fraction, $E_N$, and 
   momentum fraction of short-lived kaons that decay to charged pion 
  pairs with lab-frame decay length between 0.5 and 50 cm, $K_{S_{\pi^+ \pi^-}}$, 
  for strange- (orange) and down- (blue) quark jets arising from $Z$-boson decay. 
  The lego plots cover a total area of 1.2 $\times$ 1.2 in the $\eta-\phi$ plane 
  with pixel size $0.1 \times 0.1$. 
  Each jet image within the average is centered and rotated.  
}
 \label{fig:legoplot}
\end{figure}

A more refined picture of the observable differences between strange- and down-quark 
jets can be gleaned 
from the distributions of detector-level observables in the plane transverse 
to the jet direction. 
Lego plots  of the jet averaged charged track momentum fraction, neutral hadronic and neutral 
electromagnetic energy fractions, and $K_S \to \pi^+ \pi^-$ momentum fraction, 
$T, H_N, E_N, K_{S_{\pi^+ \pi^-}}$,
over the $\eta - \phi$ plane 
obtained from \textsc{Delphes} 
for the leading jet in the $p_T=45$ GeV event samples from $Z$-boson decay are shown in Fig.~5. 
Orange legos correspond to strange-quark jets, and blue to down-quark jets. 
The lego images cover a total area of $1.2 \times 1.2$ in the $\eta - \phi$ plane, with 
pixel of size of $0.1 \times 0.1$. 
The charged track momentum fractions are summed over individual charged tracks 
within each pixel, and the neutral energy fractions are summed over the appropriate 
calorimeter towers with centers that lie within each pixel.
Following the jet preprocessing procedure used in previous studies \cite{Macaluso:2018tck}
each individual jet image is shifted, rotated, and flipped so that the centroid of 
 $T+E_N+H_N$ coincides with the central pixel, the principal axis is in the vertical direction,  
 and the maximum intensity is in the upper right region. 
 
A number of observable differences between strange-quark and down-quark jets 
are apparent in the lego images in Fig.~5. 
The average neutral hadronic energy fraction is larger for strange-quark jets, 
and conversely the average neutral electromagnetic energy fraction is larger for down-quark jets, 
consistent with the pie charts in Fig.~1. 
Most importantly though, these differences are localized mainly in the highest intensity central pixel at the centroid  
of the jets. 
The next highest intensity pixel to the left of central pixel in the images, shows a somewhat 
reduced difference for these observables, with even smaller differences in the remaining pixels. 
This arises because after QCD showering the primary hard quark can 
carry a significant fraction of the jet total momentum, and therefore by definition be near the centroid of the jet. 
For a primary hard strange-quark that hadronizes in a long-lived neutral kaon, the resulting excess neutral hadronic energy 
then tends to be localized near the centroid of the jet.  
Likewise, for a primary hard down-quark that hadronizes in a neutral pion that then promptly decays to photon pairs, 
the resulting excess neutral electromagnetic energy tends to be localized near the centroid of the jet. 
So the greatest power of neutral hadronic and neutral electomagnetic 
energy fraction to discriminate strange- and down-quark jets is expected to come from the central region 
of jets.  

Another striking difference between strange- and down-quark jets that is apparent in the lego images 
in Fig.~5 
is the larger 
momentum fraction for strange-quark jets with $K_S \to \pi^+ \pi^-$ decays that can be reconstructed. 
Just as for the neutral hadronic energy fraction, this excess momentum fraction tends to be localized near the 
centroid of the jets.  
This difference is more pronounced than that in the neutral hadronic or neutral electromagnetic energy fractions 
near the centroid of the jets because reconstructing the leading short-lived kaon with in-flight decay 
$K_S \to \pi^+ \pi^-$ isolates a single strangeness carrying particle, which in a strange-quark jet
has a good probability of containing the primary hard strange-quark 
with the net unit of strangeness within the jet. 
However, because of the branching ratio and very conservative reconstruction efficiency assumed here, 
only 8\% of strange-quark jets and 5\% of down-quark jets have such reconstructed decays-in-flight. 
Even so, for those strange-quark jets with a reconstructed $K_S \to \pi^+ \pi^-$ near the centroid of the jet, 
this observable 
provides better discrimination from down-quark jets than does the neutral hadronic or neutral electromagnetic 
energy fractions near the centroid. 

Finally, it is worth noting that 
the charged track momentum fractions for strange- and down-quark jets 
in Fig.~5 are very similar on average everywhere in the transverse plane of the jets. 
This is because, as described in section 3.1, the probability for a down-quark to hadronize in  a 
charged pion is very similar to the probability for a strange-quark to hadronize in a 
charged kaon. 

The important features of the lego images for strange- and down-quark jets  
obtained from \textsc{Delphes} 
for the leading jet in the $p_T > 200$ GeV event samples 
are very similar to those outlined above, 
except that the overall intensity of the $K_S \to \pi^+ \pi^-$ momentum fraction 
 is smaller in both cases. 
For larger jet total momentum a higher fraction of the short-lived kaons reach the HCAL 
before decaying, resulting in relatively fewer decays-in-flight.


\section{Strange Versus Down Jet Tagging Algorithms}
\label{sec:taggers}

Some of the detector-level observables presented in the previous section 
show clear differences between strange- and down-quark jets.
In this section we describe a number of different tagging algorithms 
that exploit these detector-level differences as handles to on average distinguish 
between strange- and down-quark jets. 
The algorithms range from simple cuts on single whole-jet variables, 
BDTs with a few whole-jet variables, to 
deep learning CNNs with appropriately chosen multi-color jet images.  
The algorithms and their input sources and variables are listed in 
Table \ref{tagger-input-table}.
Results and comparisons of the performance of the 
algorithms are presented in section 5.


\renewcommand{\arraystretch}{1.5}
\begin{table}[!t]
\begin{center}
{\tabcolsep = 5mm
\begin{tabular}{ccl}
 Algorithm & Input Source & \multicolumn{1}{c}{Input Variable(s)}   \\[2ex]
Cut1~~ & \textsc{Delphes}  & $H_N - E_N $  \\
Cut1+ & \textsc{Delphes}  & $ H_N - E_N +  K_{S_{\pi^+ \pi^-}}$    \\
BDT3 & \textsc{Delphes}  & $H_N \, , \, E_N \, , \, T $\\
BDT4 & \textsc{Delphes}  & $H_N \, , \, E_N \, , \, T \, , \, K_{S_{\pi^+ \pi^-}}$  \\
CNN3 & \textsc{Delphes}  &  $H_N \, , \, E_N \, , \, T $  ~~~~~~~~~~~~~13$\times$13 Jet Image \\
CNN4 & \textsc{Delphes}   & $H_N \, , \, E_N \, , \, T \, , \, K_{S_{\pi^+ \pi^-}}$  ~13$\times$13 Jet Image \\
Truth Cut1 &  \textsc{Pythia8}  & $- \pi^0 + K_L + K_S + K_{S_{\pi^+ \pi^-}} $  \\
Truth BDT3 & \textsc{Pythia8}  & $\pi^0 \, , \, K_L \, , \, K_S +  K_{S_{\pi^+ \pi^-}}$    \\
\end{tabular}}
\end{center}
 \vspace{-0.3cm}
 \caption{
 Strange versus down jet tagging algorithms input source and variable(s) utilized in this work.  
 \textsc{Pythia8}  truth-level variables $\pi^0$ and $K_L$ are jet total momentum fractions
  in neutral pions and long-lived neutral kaons respectively, 
  $K_S$ is total momentum fraction for short-lived neutral kaons that decay 
  with a lab-frame distance greater than 150 cm, and $K_{S_{\pi^+ \pi^-}}$ is total momentum 
  fraction for short-lived neutral kaons that decay to charged pion pairs with a lab-frame 
  distance between 0.5 and 50 cm. 
  \textsc{Delphes}  variable $H_N$ is jet neutral hadronic energy fraction, $E_N$ is electromagnetic 
  energy fraction, $T$ is track momentum fraction, and $K_{S_{\pi^+ \pi^-}}$
  is   as defined above. 
  \textsc{Delphes}  variables for Cut and BDT algorithms are for the entire jet.
  For CNN algorithms the \textsc{Delphes}  variables are defined over a centered and rotated 
  13$\times$13 jet image that 
  covers 1.2$\times$1.2 in the $\eta \! -\! \phi$ plane. 
    }
 \label{tagger-input-table}
\end{table}
\renewcommand{\arraystretch}{1}


\subsection{Single Variable Cut-Based and Multi-Variable BDT Classifiers}

The simplest possible classifier for distinguishing strange-quark from down-quark jets 
is a simple cut on a single whole-jet detector based variable.  
As discussed in section 3.2, in going from a strange-quark jet to a down-quark jet, 
the neutral energy fraction, $H_N$, and 
neutral electromagnetic energy fraction, $E_N$,
are anti-correlated.  
So $H_N-E_N$ is a very simple whole-jet variable that should provide some discrimination, 
which we refer to as Cut1. 
The momentum fraction $K_{S_{\pi^+ \pi^-}}$ 
of short-lived kaons with reconstructed decays-in-flight to charged pion pairs, 
$K_S \to \pi^+ \pi^-$, provides additional discriminating power. 
So we also use the whole-jet combination $H_N - E_N + K_{S_{\pi^+ \pi^-}}$, 
which we refer to as Cut1+.
We have also considered other linear combinations of the four observables discussed in section 3.2, 
$H_N, E_N, T, K_{S_{\pi^+ \pi^-}}$, but Cut1 and Cut1+ give the best single variable discriminating power  
at the whole-jet level.  

In order to account for correlations among the four 
detector-level observables listed above, 
we utilize a BDT algorithm with whole-jet inputs of either 
$H_N, E_N, T$ 
or 
$H_N, E_N, T, K_{S_{\pi^+ \pi^-}}$, 
which we refer to as BDT3 and BDT4 respectively. 
The BDT networks are trained to distinguish between strange- and down-quark jets 
with either the $p_T = 45$ GeV or $p_T > 200$ GeV 
event samples described in section 2. 
The algorithms are tested on independent event samples, with the 
BDT 
output used as a single classifier variable on which to cut.

It is instructive to compare these detector-level whole-jet classifiers  
with classifiers based on analogous truth-level whole-jet information. 
For a single truth-level variable we employ the whole-jet neutral kaon momentum fraction 
minus the neutral pion momentum fraction discussed in section 3.2. 
To retain some irreducible detector level effects, only short-lived kaons that reach 
the HCAL at 150 cm from the beam axis, or decay to charged pion pairs with a lab-frame 
distance between 0.5 and 50 cm are included. 
We refer to this classifier as Truth Cut1. 
For a classifier that accounts for correlations among truth-level variables 
we utilize a BDT algorithm with whole-jet inputs of the neutral pion momentum fraction, 
the long-lived neutral kaon momentum fraction, and the short-lived neutral kaon momentum fraction 
with the restrictions above. 
We refer to this classifier as Truth BDT3.  
We view this classifier as 
roughly 
setting the maximal performance that can be achieved for 
discriminating strange-quark from down-quark jets 
with whole-jet variables, 
given some of 
the irreducible detector-level effects of the current generation of general purpose collider detectors.


\subsection{Deep Learning CNN-Based Jet Image Classifiers}

\begin{figure}[!t]
  \begin{center}
  \vspace{0.5cm}
   \includegraphics[clip, width=16.75cm]{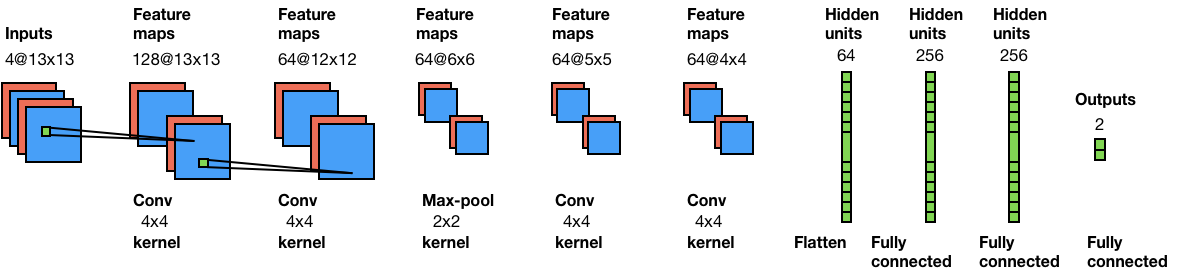}
  \end{center}
  \caption{The deep learning convolutional neural network  
  architecture CNN4 utilized here for discriminating 
  strange-quark and down-quark jets. 
  The inputs are 13$\times$13 jet images 
  from the \textsc{Delphes} detector simulation 
  covering a total area of 
  1.2$\times$1.2 in the $\eta \! - \! \phi$ plane 
   with 4 colors corresponding 
  to the normalized variables $H_N , E_N, T, K_{S_{\pi^+ \pi^-}}$ in each pixel.   
  The CNN3 architecture is identical with 3 colors corresponding to 
  $H_N , E_N, T$ in each pixel.   
  The images are centered and rotated.  
  Two outputs correspond to strange-quark or down-quark jet. 
  }
  \label{fig:CNN}
\end{figure}

Going beyond whole-jet variables, a more detailed picture of the
differences 
between strange- and down-quark jets is provided by the distribution of detector-level 
observables in the plane transverse to the jet direction, 
as discussed in section 3.2. 
In order to capture these finer details we use a CNN algorithm  
with pixelated multi-color jet image inputs of either the detector-level observables
$H_N, E_N, T$ 
or 
$H_N, E_N, T, K_{S_{\pi^+ \pi^-}}$, 
which we refer to as CNN3 and CNN4 respectively. 
The jet image inputs for each of these detector-level observables correspond precisely 
to those shown in Fig.~5, with $13 \times 13$ pixels of size $0.1 \times 0.1$ in the $\eta - \phi$ plane 
with the same
preprocessing described in section 3.2. 

The architecture of the CNN algorithm we utilize is shown in Fig.~\ref{fig:CNN}, 
and is a variation of one used in 
previous studies \cite{Macaluso:2018tck}.
The 3 or 4 input images from each jet
enter the first convolutional layer of 128 filters through a local $4 \times 4$ kernel.
The generated 128 feature maps are then coupled to the next convolutional layer of 64 filters with another  
local $4 \times 4$ kernel.
This is followed by a max-pooling layer with a $2 \times 2$ reduction factor.  
The resulting 64 feature maps are coupled sequentially through two more convolutional layers, 
each of which has 64 filters  with a local $4 \times 4$ kernel.
These 64 feature maps are then coupled directly through three sequential fully-connected 
layers with 64, 256 and 256 neurons respectively.
Finally, the output layer has two neurons corresponding to strange- or down-quark jets. 
A softmax function is used as activation to output the classification probabilities. 
All the convolutional layers and fully-connected layers in the CNN 
have Rectified Linear Units (ReLUs) as activation.
The CNNs are trained to distinguish between strange- and down-quark jets 
with either the $p_T = 45$ GeV or $p_T > 200$ GeV 
event samples described in section 2. 
Independent event samples are employed for training and testing.  
We use cross entropy for the loss function and Adadelta \cite{adadelta} for the optimizer.
The CNN output is used as a single classifier variable on which to cut.  

We have also tried Adam \cite{adam} for the CNN optimizer but the classification
results do not show any improvement.
In addition, 
we have changed the number of filters or added another max-pooling layer just before the fully-connected layers, 
again with no significant improvement. 
Finally, we have tried ResNet algorithms along the lines of another previous study 
\cite{Qu:2019gqs} but they did not offer superior performance over the CNN architecture.


\section{Strange Versus Down Jet Tagging Algorithms Performance}
\label{sec:deep}

All of the algorithms outlined in the last section provide some level of discrimination between 
strange- and down-quark jets.  
In the first sub-section below, the training sample sizes required for the BDT and CNN algorithms 
are discussed. 
In the next sub-section the relative and absolute performance of all the algorithms are compared. 
In the final sub-section the correlation between the classifier output of one of the CNN algorithms 
and the long-lived kaon momentum fraction is investigated.


\subsection{Training Sample Size for BDT and CNN Algorithms}

\begin{figure}[!t]
  \begin{center}
  \vspace{0.5cm}
   \includegraphics[clip, width=10cm]{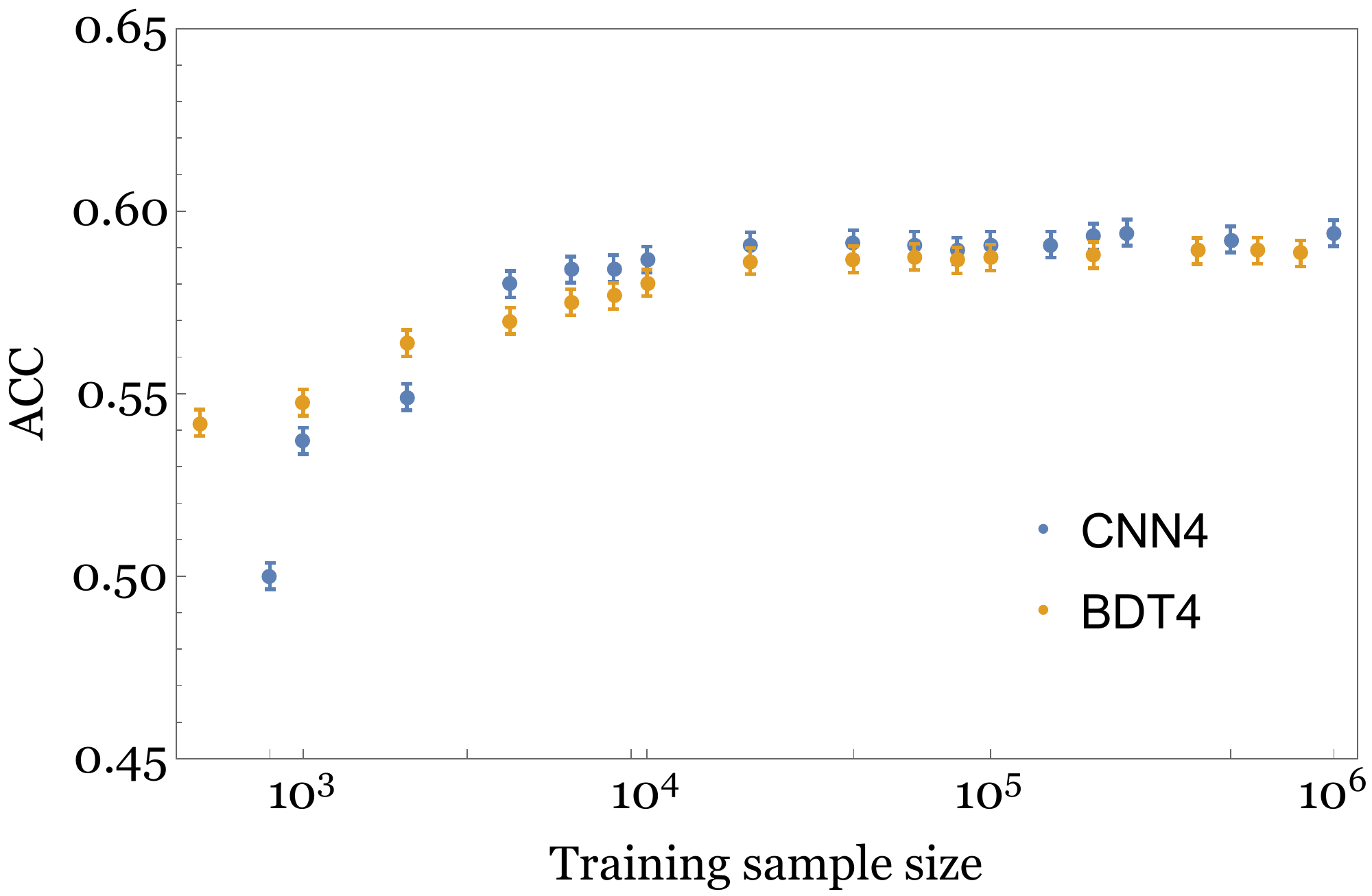}
  \end{center}
  \caption{
  Classification accuracy of strange-quark versus down-quark central jets from $Z$-boson decay 
   for the BDT4 (orange) and CNN4 (blue) strange jet tagging algorithms, 
  as a function 
  of the training sample size for a fixed independent testing sample of $10^4 +10^4$ strange 
  plus down jets. 
  Error bars correspond to binomial statistical standard deviation of the testing sample. 
  The BDT4 and CNN4 
  tagging algorithm inputs are given in Table 1.
 }
  \label{fig:SampleSize}
\end{figure}

The BDT and CNN algorithms are trained to distinguish between
strange- and down-quark 
jets with either the $p_T = 45$ GeV or $p_T > 200$ GeV event samples described in section 2. 
An important figure of merit for these algorithms 
is the number of training events required to obtain good performance. 
As an overall measure of the performance for a given number of training events, 
we use the best accuracy, ACC, 
defined to be the average of the true positive and true negative rate, 
maximized over all values of the classifier output variable. 
The best accuracy for the BDT4 and CNN4 algorithms are shown in 
Fig.~\ref{fig:SampleSize} as a function of the training event sample size for 
the $p_T=45$ GeV event samples from $Z$-boson decay. 
For both algorithms the performance saturates with more than roughly $10^4$
training events. 
By contrast, at least 10 times more training events were needed in a previous study to 
saturate performance for boosted hadronic top-quark tagging \cite{Macaluso:2018tck}. 
This is perhaps not surprising since boosted hadronic top-quark jets have more 
localized substructure in the plane transverse to the jet direction than do 
strange- and down-quark jets.


\subsection{Comparison of Strange Jet Tagging Algorithms }

\begin{figure}[!t]
  \begin{minipage}{0.48\hsize}
  \begin{center}
   \includegraphics[clip, width=7.5cm]{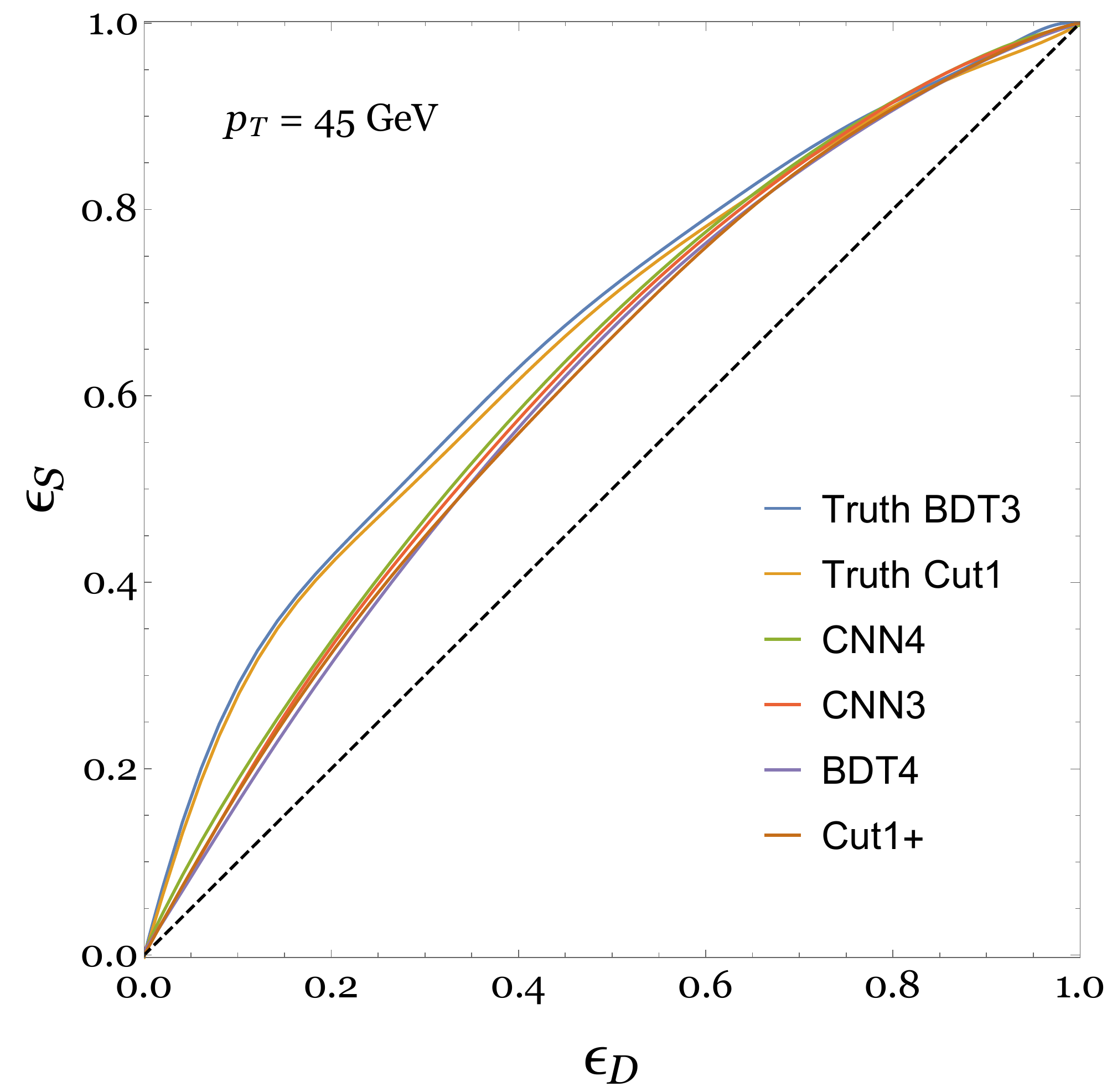}
  \end{center}
 \end{minipage}
 \hspace{0.5cm}
 \begin{minipage}{0.48\hsize}
  \begin{center}
   \includegraphics[clip, width=7.5cm]{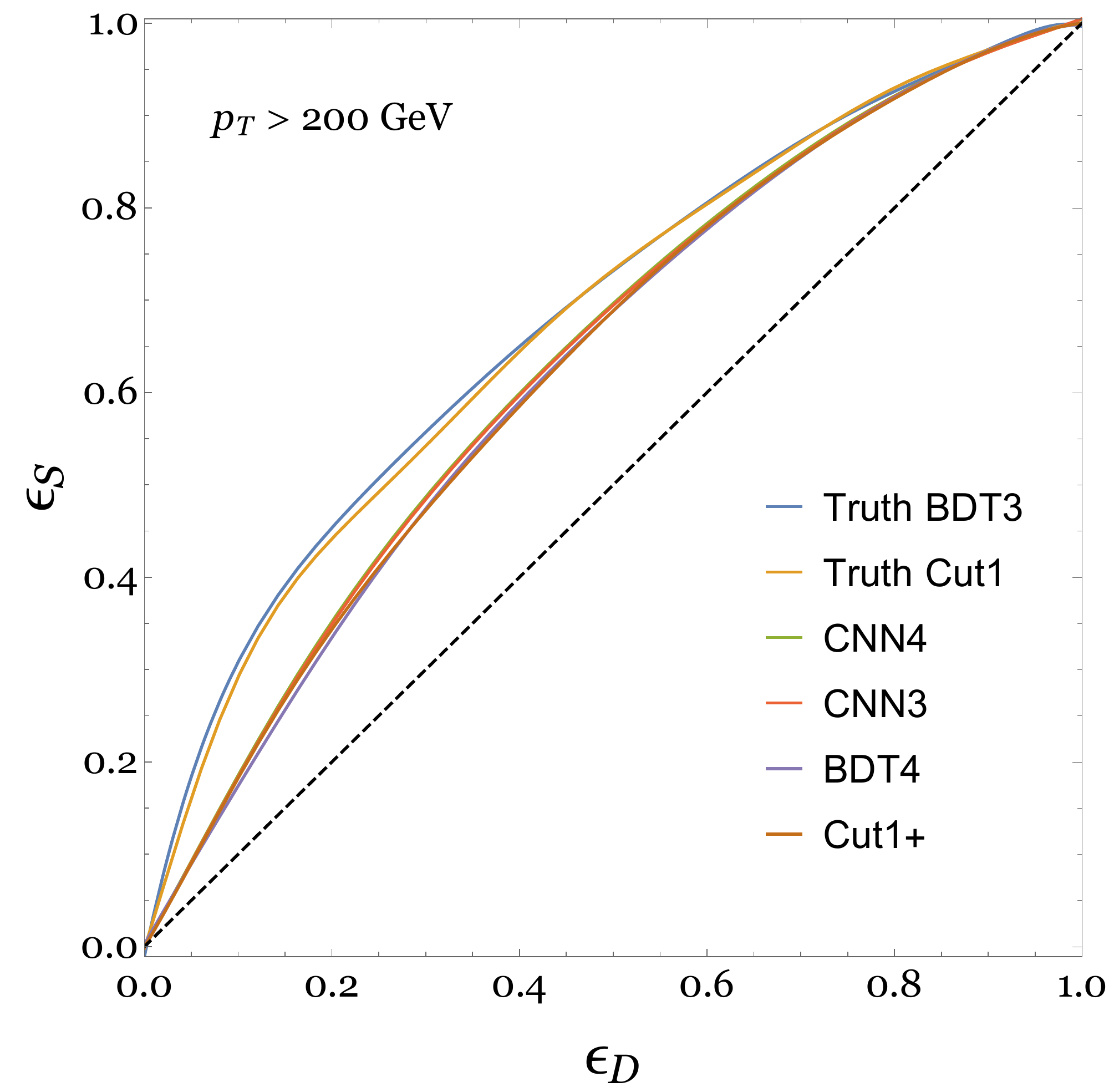}
  \end{center}
 \end{minipage}
 \vspace{0.3cm}
  \caption{
  Strange jet tagging algorithm 
  ROC curves 
  for the fraction of strange-quark central jets correctly tagged as strange jets, $\epsilon_S$, 
   as a function of the 
  fraction of down-quark central jets incorrectly tagged as strange jets, $\epsilon_D$, 
  arising from either $Z$-boson decay or QCD initiated 13 TeV proton-proton collisions with jet 
  $p_T > 200$ GeV.  
  Tagging algorithm inputs are given in Table 1.
  The dashed diagonal line corresponds to 
  random tagging of strange- versus down-quark jets. 
  }
\label{fig:ROC}
  \end{figure}

\begin{figure}[!t]
  \begin{minipage}{0.48\hsize}
  \begin{center}
   \includegraphics[clip, width=7.5cm]{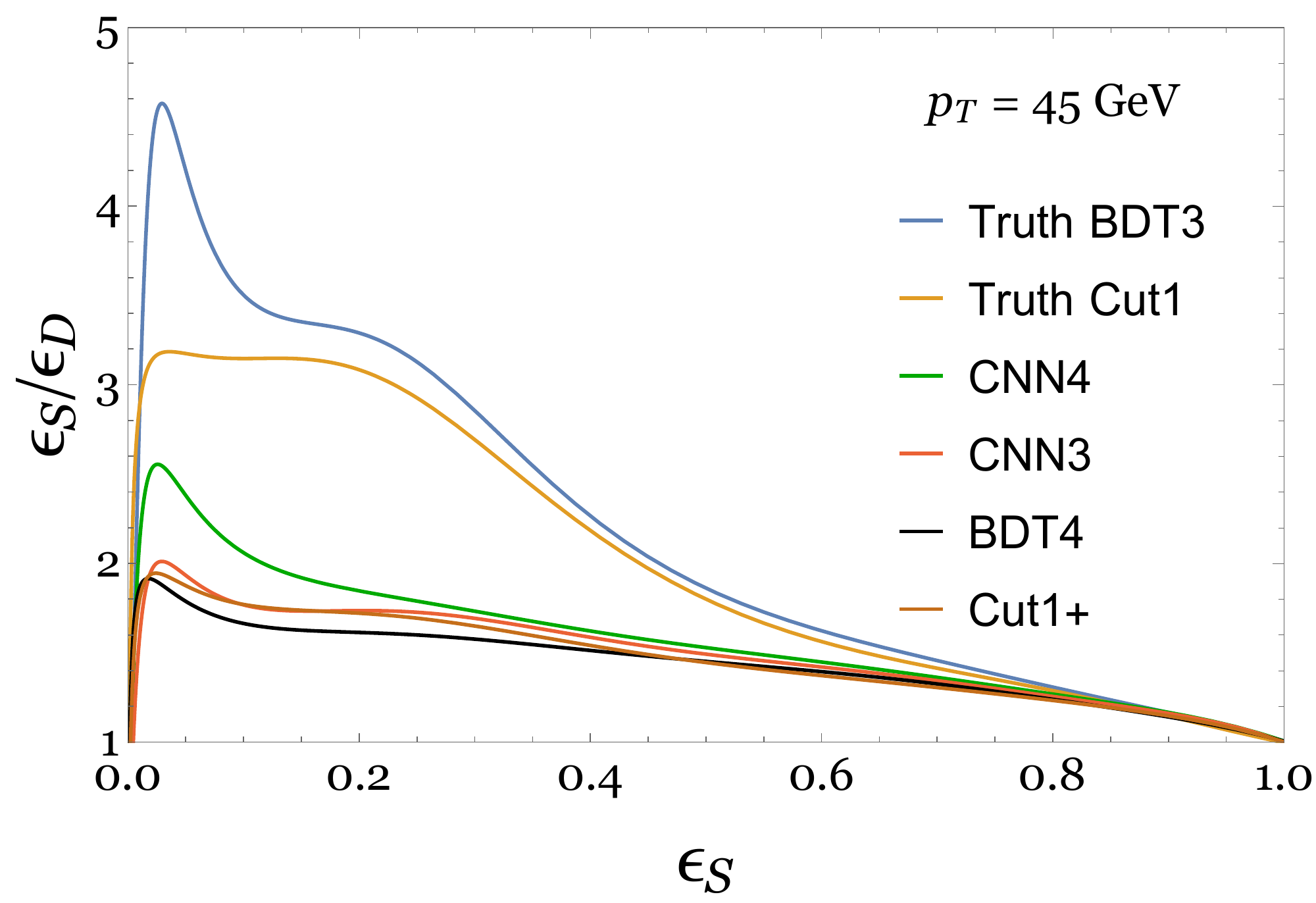}
  \end{center}
 \end{minipage}
 \hspace{0.5cm}
 \begin{minipage}{0.48\hsize}
  \begin{center}
   \includegraphics[clip, width=7.5cm]{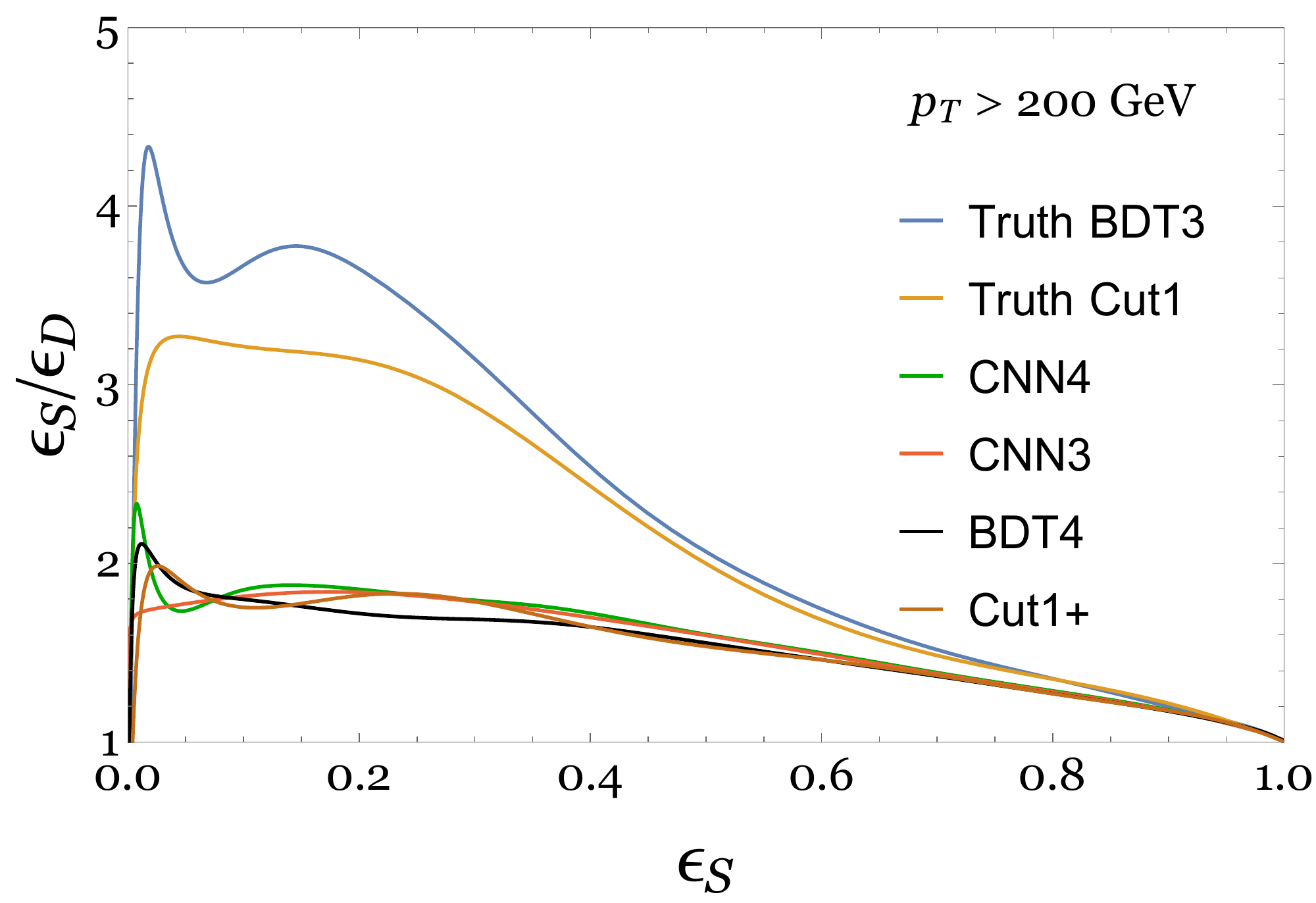}
  \end{center}
 \end{minipage}
 \vspace{0.5cm}
  \caption{  The ratio of strange-quark to down-quark central jets tagged as strange jets, 
  $\epsilon_S / \epsilon_D$, 
   as a function of the 
  fraction of strange-quark jets tagged as strange jets, $\epsilon_S$, 
  arising from either $Z$-boson decay or QCD initiated 13 TeV proton-proton collisions with jet 
  $p_T > 200$ GeV.  
  The tagging algorithm inputs are given in Table 1. 
}
\label{fig:Ratio}
  \end{figure}

All the strange- versus down-quark tagging algorithms presented in section 4
output a single classifier variable. 
Jets within events in a testing 
set with classifier values larger than a threshold or cut value may be 
tagged as strange-quark like jets. 
With this prescription, it is convenient to define 
the tagging rate, $\epsilon_S$, to be the fraction of strange-quark jets 
in a testing sample set that are tagged as strange-quark like jets, 
and the mistag rate, $\epsilon_D$, to be the fraction of down-quark jets in a testing 
sample set that are tagged as strange-quark like jets.  
The Receiver Operating Characteristic (ROC) curve for a tagging algorithm 
is then defined as the pair $\epsilon_S, \epsilon_D$ as a function of the classifier 
cut value. 

The ROC curves for the whole-jet detector-level algorithms Cut1+ and BDT4, 
the whole-jet truth-level algorithms Truth Cut1 and Truth BDT3, 
and the detector-level jet image algorithms CNN3 and CNN4 are shown in Fig.~\ref{fig:ROC}
for both the $p_T = 45$ GeV and $p_T > 200$ GeV event samples described in section 2. 
For comparison, a 
random tagging of strange- versus down-quark jets would yield a diagonal ROC curve.  
The truth-level whole-jet algorithms have better performance than any of the detector-level 
algorithms for nearly all classifier cut values. 
This indicates that the detector-level observables tend to 
wash out the information in the more idealized truth-level variables. Of particular importance in this regard, 
the discriminating power from the larger momentum weighted fraction of 
long- and short-lived neutral kaons that reach the HCAL in strange-quark jets is 
partially overcome by the larger momentum weighted fraction of neutrons 
in down-quark jets, as discussed in section 3.1. 

In order to better expose the differences between the strange-quark jet tagging algorithms, 
ratio ROC curves, defined as the pair $\epsilon_S / \epsilon_D , \epsilon_S$ 
as a function of the classifier cut value, 
are shown in Fig.~\ref{fig:Ratio}. 
For the $p_T = 45$ GeV jet event samples from $Z$-boson decay, 
among the algorithms based on detector-level observables, CNN4 
with jet image inputs 
shows the best discrimination between strange- and down-quark jets 
for all classifier cut values. 
It is better than the whole-jet BDT4 algorithm with the same 
detector-level inputs integrated over the plane transverse to the jet direction. 
As discussed in section 3.2, 
the differences between strange- and down-quark jets 
in hadronic neutral energy, electromagnetic neutral energy, and short-lived kaon $K_S \to \pi^+ \pi^-$ reconstructed momenta, 
are localized mainly near the centers of jets. 
The spatial  distribution of these detector-level observables in the transverse plane is available to the 
CNN4 algorithm, but not the BDT4 algorithm. 

The CNN3 algorithm with detector-level inputs is essentially identical in architecture to the 
CNN4 algorithm, except that it doesn't include the 
momentum fraction of 
reconstructed short-lived kaon decays-in-flight to charged pion pairs, $K_S \to \pi^+ \pi^-$. 
This accounts for the better performance of CNN4 compared with CNN3 for 
the $p_T = 45$ GeV jet event samples shown in Fig.~\ref{fig:Ratio}. 
The difference in performance is more pronounced 
for smaller values of the strange-quark jet tagging rate, $\epsilon_S$.  As discussed in section 3.2, because of the branching ratio and very conservative reconstruction 
efficiency only 8\% of strange-quark jets in this event sample have reconstructed $K_S \to \pi^+ \pi^-$ 
decays-in-flight. 
For smaller values of the strange-quark jet tagging rate, obtained for harder cuts on the 
classifier output variable, 
the CNN4 algorithm makes effective use of the somewhat rare
reconstructed $K_S \to \pi^+ \pi^-$ decays-in-flight. 
With this additional discriminating handle, the CNN4 algorithm obtains peak 
performance of $\epsilon_S / \epsilon_D \simeq 2.5$ 
in distinguishing strange- and down-quark jets in this event sample at the somewhat small 
value of strange-quark tagging rate of $\epsilon_S \simeq 0.02$.  
 
The whole-jet truth-level Truth BDT3 algorithm has better performance than the whole-jet single variable 
truth-level Truth Cut1 algorithm  for all values of the classifier variable cut. 
The difference is more pronounced for smaller values of the strange-quark jet tagging rate. 
Just as above, this arises mainly from the reconstructed $K_S \to \pi^+ \pi^-$ decays-in-flight. 
The whole-jet Truth BDT3 algorithm makes use of these somewhat rare reconstructed decays-in-flight 
at small values of the strange-quark jet tagging rate. 
The occurrence of such decays-in-flight is infrequent enough that the information about 
the spatial distribution is less important than for other observables. 
In contrast, for Truth Cut1 
the contribution of the momentum fraction of these decays-in-flight 
to the single whole-jet variable of this algorithm is diluted. 
The whole-jet detector-level BDT4 algorithm also makes use of the reconstructed  
 $K_S \to \pi^+ \pi^-$ decays-in-flight to obtain better performance at small values of the 
strange-quark jet tagging rate. 

The detector-level whole-jet algorithms BDT4 and Cut1+ and jet image algorithms CNN3 and CNN4 
all have somewhat similar performance for nearly all classifier cut values for the 
$p_T > 200$ GeV jet event samples shown in Fig.~\ref{fig:Ratio}. 
This is unlike the $p_T = 45$ GeV jet samples for which the CNN4 jet image algorithm has noticeably 
better performance than the other detector-level algorithms for small classifier cut values. 
As discussed in section 3.1, the main observable difference between these jet samples is that 
for $p_T > 200$ GeV a sizable 
fraction of the short-lived neutral kaons reach the HCAL and deposit energy there. 
This leaves a smaller fraction of short-lived neutral kaons with decays-in-flight to charged pion pairs, 
$K_S \to \pi^+ \pi^-$, that can be reconstructed for $p_T > 200$ GeV jets. 
The primary observable handle available to distinguish strange- from down-quark jets that remains 
is then the hadronic neutral and electromagnetic neutral energy fractions, with all the detector-level 
algorithms giving somewhat similar performance. 
However, at very small values of classifier cut values, the CNN4 and BDT4 algorithms as well as Cut1+  
that all include the reconstructed $K_S \to \pi^+ \pi^-$ momentum fraction, do show some improvement in performance 
compared with the CNN3 algorithm that does not include this information. 
All of this illustrates the added benefit of including reconstructed decays-in-flight of short-lived kaons to charged pion pairs, 
$K_S \to \pi^+ \pi^-$, in strange-quark jet tagging, particularly for lower momentum jets.


\renewcommand{\arraystretch}{1.5}
\begin{table}[!t]
\begin{center}
{\tabcolsep = 5mm
\begin{tabular}{ccccccc}
 & AUC & ACC & R10  & R50  \\ 
Truth Cut1 & 0.65 (0.68) & 0.61 (0.62) & 31.9 (32.1) & 3.6  (3.9) \\
Truth BDT3  & 0.67 (0.68) & 0.62 (0.62) & 37.3 (37.1) & 3.7  (4.0)\\
Cut1~~  & 0.61 (0.63) & 0.57 (0.59) & 16.4 (17.9) & 2.7  (3.0) \\
Cut1+  & 0.62 (0.63) & 0.58 (0.60) & 17.9 (18.8)& 2.9 (3.1)   \\
BDT3 & 0.61 (0.63) & 0.59 (0.60) & 16.0 (17.1) & 2.9 (3.1) \\
BDT4 & 0.63 (0.63) & 0.60 (0.60) & 22.5 (16.6)& 3.2 (3.2) \\
CNN3 & 0.62 (0.63) & 0.59  (0.60) & 17.9 (18.4) & 3.0 (3.2) \\
CNN4   & 0.64 (0.64) & 0.60 (0.60) & 23.9 (18.8) & 3.3 (3.2) \\
\end{tabular}}
\end{center}
 \vspace{-0.3cm}
 \caption{
 Area under ROC curve, accuracy, and rejection factors 
 R10 = $1/\epsilon_D$ for $\epsilon_S = 0.1$, and 
 R50 = $1/\epsilon_D$ for $\epsilon_S = 0.5$, 
 for various strange jet tagging algorithms applied to Monte Carlo samples 
 with equal numbers of strange-quark and down-quark jets. 
 Numbers without parentheses are for strange- and down-quark 
 central jets from $Z$-boson decay, and with parentheses likewise from 
 QCD initiated 13 TeV proton-proton collisions with jet $p_T > 200$ GeV. 
}
 \label{performancetab}
\end{table}
\renewcommand{\arraystretch}{1}

Numerical measures of the performance of all the strange- versus down-quark jet tagging algorithms listed 
in Table 1 are presented in Table~\ref{performancetab} for both the $p_T=45$ GeV and $p_T > 200$ GeV 
event samples described in section 2.  
Among the detector-level algorithms, the CNN4 algorithm with jet image inputs has the best performance. 
The area under the ROC curve, AUC, for this algorithm is 0.64 for both event samples.  
For comparison, a perfect tagging algorithm would have AUC equal to 1.0 and a random algorithm would have 0.5. 
The accuracy, ACC defined in section 5.1, for this algorithm is 0.60.  
The down-quark jet rejection factor for the CNN4 algorithm, defined to be $1/ \epsilon_D$, 
reaches a value of 23.9 for $\epsilon_S = 0.1$, corresponding to $\epsilon_S / \epsilon_D = 2.39$, 
for the $p_T = 45$ GeV event sample. 

 \begin{figure}[!t]
  \begin{minipage}{0.48\hsize}
  \begin{center}
   \includegraphics[clip, width=7.5cm]{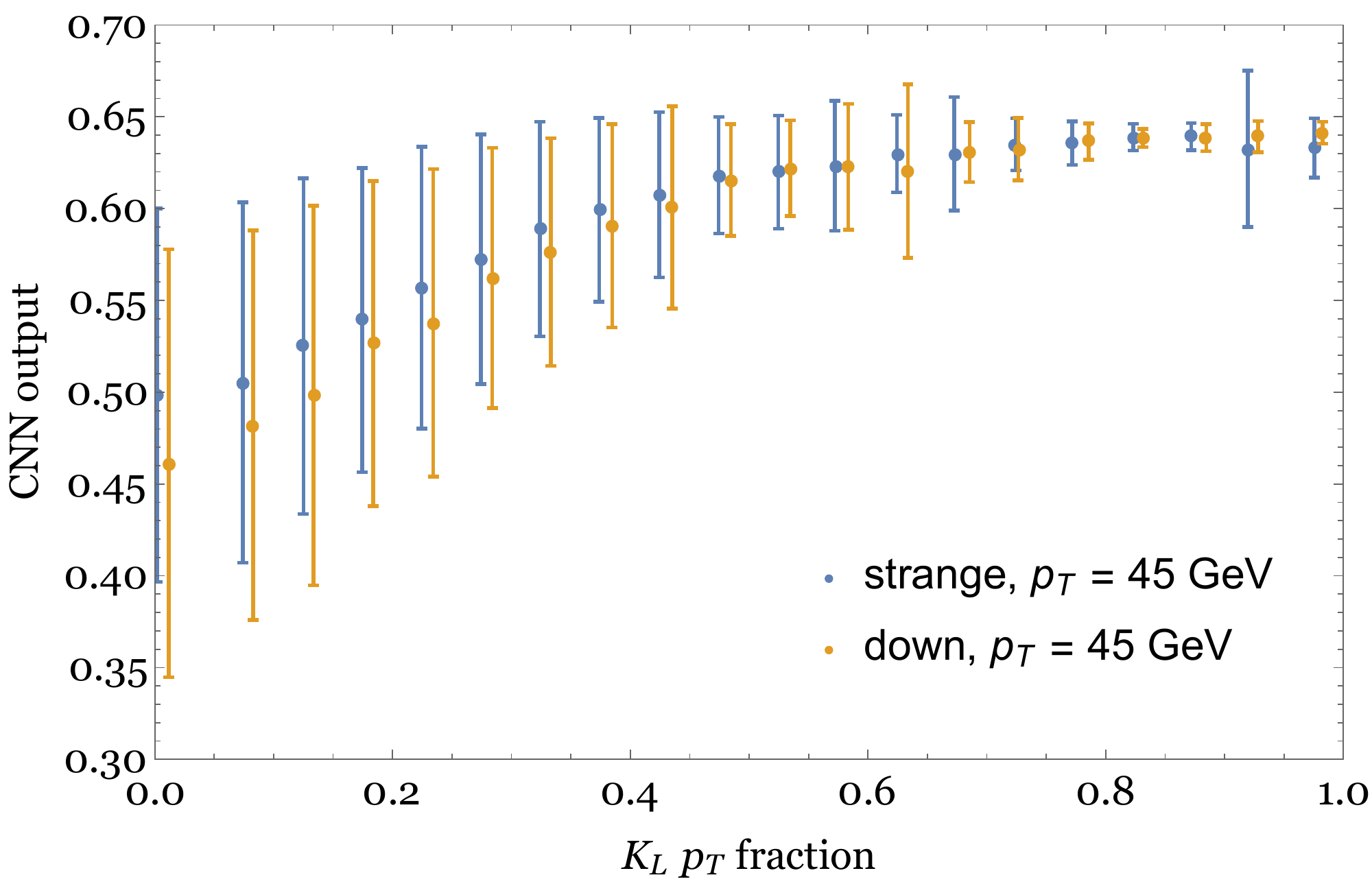}
  \end{center}
 \end{minipage}
 \hspace{0.5cm}
 \begin{minipage}{0.48\hsize}
  \begin{center}
   \includegraphics[clip, width=7.5cm]{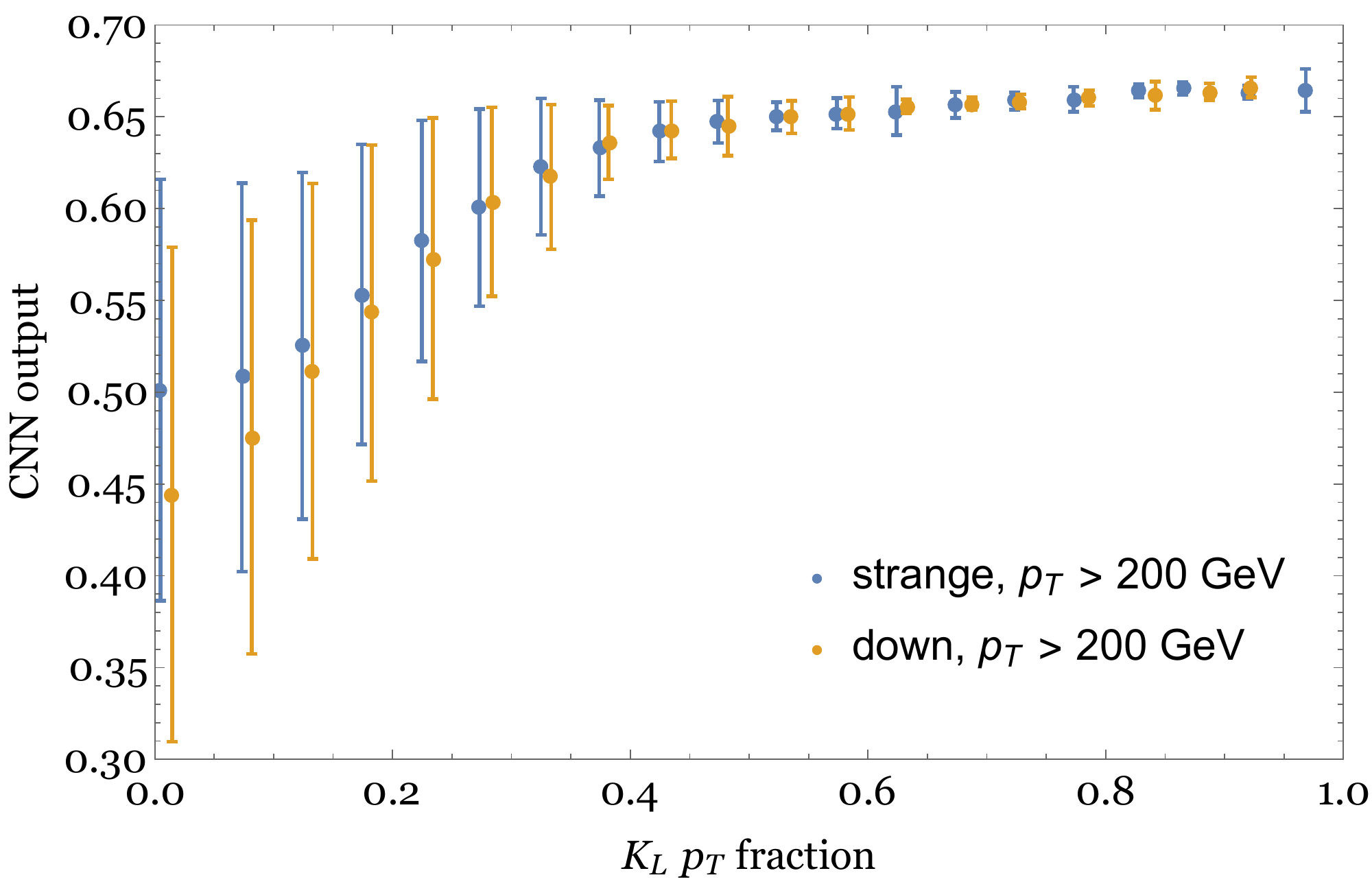}
  \end{center}
 \end{minipage}
 \vspace{0.3cm}
  \caption{
  CNN3 classifier variable output as a function of \textsc{Pythia8} truth-level long-lived kaon 
  transverse momentum fraction for strange- or down-quark central jets 
  arising from either $Z$-boson decay or QCD initiated 13 TeV proton-proton collisions with 
  jet $p_T > 200$ GeV.  The CNN3 inputs are given in Table 1. 
  The training and independent  testing samples are $10^6 + 10^6$ and $10^5 + 10^5$ 
  events respectively of strange- plus down-quark jets.
  The central dot in each bin corresponds to the average CNN3 classifier variable output, 
  and the error bar corresponds to the variance in each bin. 
   }
  \label{fig:NNoutput}
\end{figure}


\subsection{Correlation of CNN3 Algorithm with Long-Lived Kaon Momentum Fraction}

In a general purpose high energy collider detector, 
the primary observable differences between strange- and down-quark jets 
are for the momentum fraction of neutral kaons
with a concomitant anti-correlated momentum fraction of neutral pions, 
as detailed in section 3.
In order to assess quantitatively the extent to which a detector-level 
strange-quark jet tagging algorithm 
is exploiting these primary physical differences, it is useful to 
look at the correlation between the classifier output variable and the neutral kaon 
momentum fraction. 
To expose this correlation it is most instructive to consider the CNN3 algorithm 
with detector-level jet image inputs of only the 
hadronic neutral and electromagnetic neutral energy fractions, and the track momentum fraction, 
$H_N, E_N, T$. 
This algorithm does not include reconstructed decays-in-flight of short-lived neutral kaons, 
so the relevant neutral kaon momentum fraction with which to compare is that of the 
long-lived neutral kaons that reach the HCAL and contribute to the $H_N$ detector-level observable. 

The average value and variance of the classifier output variable for the CNN3 algorithm 
for strange- and down-quark jets in both the $p_T=45$ GeV and $p_T > 200$ GeV event samples 
are shown in Fig.~\ref{fig:NNoutput} in bins of the truth-level 
long-lived neutral kaon momentum fraction 
(anomalously large variance results in a few cases because of low 
statistics). 
There is a clear correlation between the classifier output variable and the long-lived neutral kaon momentum 
fraction, with a smaller variance for larger values of the momentum fraction. 
At large values of the long-lived neutral kaon momentum, the classifier output variable saturates 
to a unique value (with very small variance) that depends only on the jet transverse momentum. The correlation and saturation occurs for both the strange- and down-quark jet samples.  
This indicates that, through the network training, the CNN3 algorithm has learned to classify a 
jet as strange-quark like based largely on the long-lived neutral kaon momentum fraction, 
particularly when it is large.  

For smaller values of the long-lived neutral kaon momentum fraction the 
average value of the CNN3 classifier output variable shown in 
Fig.~\ref{fig:NNoutput} is different for strange- and down-quark jets, but with a large variance. 
This indicates that the CNN3 algorithm has learned to use other information to distinguish
strange- from down-quark jets when the long-lived neutral momentum fraction is small. 
For these cases there is a good probability that another type of particle carries the net unit of strange-ness 
in a strange-quark jet, such as a short-lived neutral kaon that reaches the HCAL and contributes 
to the $H_N$ detector-level observable, or a short-lived kaon that decays-in-flight to a pair of either 
neutral or charged pions that can contribute to the $E_N$ or $T$ detector-level observables respectively, 
depending on the lab-frame decay distance. 
This is also why the CNN3 classifier output variable saturates in the $p_T > 200$ GeV jet samples
for smaller values of the long-lived neutral kaon momentum fraction 
 than 
in the $p_T = 45 $ GeV jet samples.  
In the former case, more short-lived neutral kaons reach the HCAL and contribute to the $H_N$ detector-level 
variable than in the latter case, as discussed in section 3.1.


\section{Possible Application of Strange-Quark Jet Tagging}  

The ultimate utility of any object reconstruction and 
tagging algorithm is its deployment 
in the analysis of data from a high energy collider detector. 
In the first two sub-sections below, 
the statistical improvement obtained from a general binary tagging algorithm in a new physics search, or measurement of known process, is presented. In the final sub-section we discuss 
the decay of a $W$-boson to a charm-quark and either a strange- or down-quark, 
as a high purity source of strange-quark jets.


\subsection{Improvements to New Physics Searches from Tagging Algorithms} 

The classifier output variable from a tagging algorithm may be treated 
like any other variable in an analysis of high energy collider data. 
For a binary tagging algorithm, two categories at either an object or event level 
may be formed based on whether the 
classifier output variable is larger or smaller than some cut or threshold value.
The resulting categories can be integrated in to and improve an analysis. 

In order to assess the impact of using the classifier output variable from a 
tagging algorithm, consider first a search for a new physics process 
in some region of the data from a high energy collider detector that, without use of the 
tagging algorithm, yields $S$  signal events and $B$ background events.
Neglecting any systematic uncertainty in the expected number of background events, and 
in the limit  $1 \ll S \ll B$ considered for simplicity here and below in this sub-section, 
the event count statistical significance for a search limit or discovery 
without the tagging algorithm 
is ${\cal S} \simeq S / \sqrt{B}$.
 
In theory-level studies of the impact of tagging algorithms 
on new physics searches, 
it is common to simply throw away events 
with classifier output variables smaller than some cut value,
and only keep the remaining events in the analysis. 
In this case a fraction $\epsilon_S$ of the signal events, 
and a fraction $\epsilon_B$ of the background events, 
remain after the classifier value cut. 
With this prescription,  the resulting event count statistical significance for a search limit or discovery
is 
\begin{equation}
{\cal S}_{\rm cut} \simeq { \epsilon_S \over \sqrt{ \epsilon_B} } \, {S \over  \sqrt{B} } 
\end{equation}
In the theory literature the quantity 
${\cal S}_{\rm cut}  / {\cal S} = \epsilon_S / \sqrt{ \epsilon_B}$ is often referred to as the statistical 
significance improvement. 
Note that this simple cut based implementation of a tagging algorithm 
with a single category, 
may or may not provide an improvement in the statistical significance 
depending on whether or not the working point defined by the pair 
$\epsilon_S, \epsilon_B$, that results from a given value of the classifier cut variable,
yields a significance improvement factor greater or less than unity respectively.

In a modern analysis of actual collider detector data, it has become standard 
to bin rather than cut in any number of variables, including tagging algorithm
classifier variables. 
With a binary tagging algorithm, binning makes use of both tagging categories, and by definition 
improves the sensitivity. A fraction $\epsilon_S$ of the signal events, 
and $\epsilon_B$ of the background events form one category, 
and a fraction $1- \epsilon_S$ of the signal events and $1- \epsilon_B$ 
of the background events form a second category. 
With this, the event count combined statistical significance for a search limit or discovery 
is the quadrature sum of the statistical significances of the two categories 
\begin{equation}
\begin{split}
{\cal S}_{\rm bin} 
\simeq  \sqrt{ \bigg(  {\epsilon_S\over \sqrt{\epsilon_B}} {S\over\sqrt{B}} \bigg)^2 +  
    \bigg(  {(1-\epsilon_S)\over \sqrt{(1-\epsilon_B)}} {S\over\sqrt{B}}  \bigg)^2 } 
= \sqrt{ 1+ {(\epsilon_S-\epsilon_B)^2\over \epsilon_B(1-\epsilon_B)}}   \,  
     {S\over\sqrt{B}} 
\label{twocategory}
\end{split}
\end{equation}
For any binary tagging algorithm working point defined by the pair 
$\epsilon_S, \epsilon_B$ this two category event count statistical significance is greater than 
that obtained from 
the single category cut based approach, or from that without use of the tagging algorithm. 
Ignoring any issues of systematic uncertainties, it is always better to bin rather than cut, or 
to not bin at all.

\begin{figure}[!t]
  \begin{center}
  \vspace{0.5cm}
   \includegraphics[clip, width=7cm]{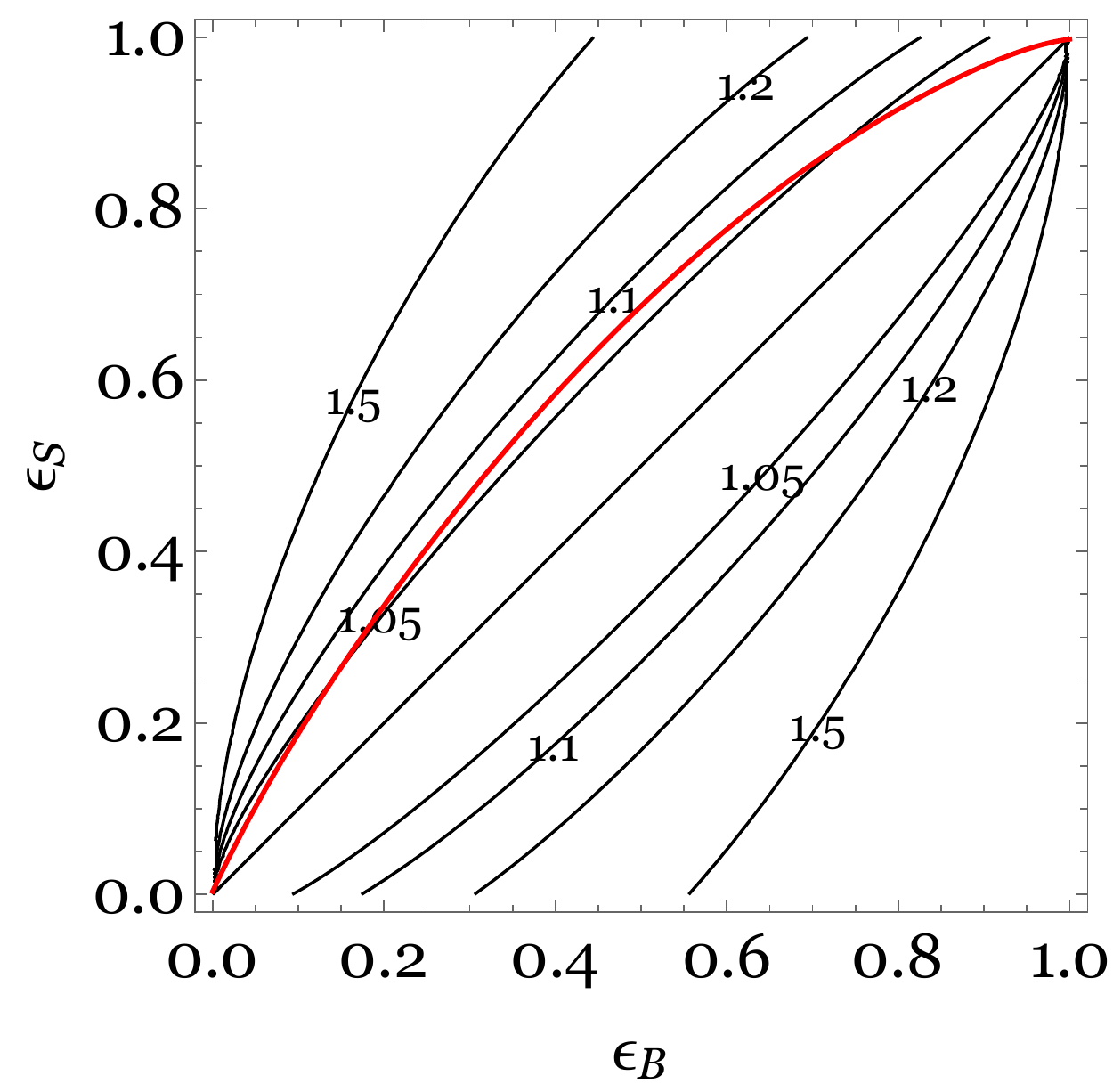}
  \end{center}
  \caption{Contours of the event count combined statistical significance improvement provided by a 
   two-category event-level tagging algorithm as a function of signal and background tagging efficiencies 
   $\epsilon_S, \epsilon_B$. 
   For reference, the ROC curve  for the CNN4 strange- versus down-quark jet tagging algorithm 
   for the $p_T = 45$ GeV jet event samples is overlaid in red. 
   }
  \label{fig:Significance}
\end{figure}

The binned event count statistical significance improvement, 
${\cal S}_{\rm bin}  / {\cal S}$,
is a useful figure of 
merit for the improvement provided by a binary tagging algorithm. 
Contours of this improvement factor are shown in Fig.~\ref{fig:Significance}
as a function of $\epsilon_S , \epsilon_B$ for a general event-level binary tagging algorithm. 
For reference, the ROC curve for the CNN4 strange- versus down-quark 
jet tagging algorithm  shown in Fig.~\ref{fig:ROC} 
for the $p_T = 45$ GeV events samples
is overlaid in Fig.~\ref{fig:Significance}.
This illustrates the statistical event count significance improvement that could be achieved with 
a binned two-category strange- versus down-quark CNN4 tagging algorithm in 
a hypothetical new physics search with signal events that include only 
a single strange-quark jet, 
with a much larger number of 
analogous  background events that include only a single down-quark jet. 
For such an idealized new physics search, the CNN4 tagging algorithm could provide a modest 
5\% improvement in the event count statistical significance over a wide range 
of classifier category values.  


\subsection{Measurements with Tagging Algorithms} 

Tagging algorithms can also be 
useful in measurements of known processes at high energy colliders. 
Just as for a new physics search, with a binary tagging algorithm, two
categories at either the object or event level may be formed based on 
a classifier output value. 
The discriminating power this offers can 
improve a measurement 
or even make feasible a measurement that was otherwise not possible.  

In order to illustrate the use of
 a tagging algorithm in a measurement 
of some known physics process, 
consider a region of the data from a high energy collider detector 
with $S$ events of one type and $B$ events of a second type. 
A binary tagging algorithm designed to distinguish these two types of events 
may be used to form two categories of events based on a cut or threshold value 
of the output classifier variable.  
Following the notation in the previous sub-section, 
the number of events in the first category is 
$N_1 = \epsilon_S \, S + \epsilon_B \, B$, and 
the number in the second category is 
$N_2 =  (1 - \epsilon_S) \, S + (1 - \epsilon_B) \, B$. 
At a given working point for the tagging algorithm defined by the pair 
$\epsilon_S , \epsilon_B$ the fraction of events of the first type in the data 
can be inferred from the  measured number of events $N_1, N_2$ in the two tagged
categories by inverting the expressions above 
\beq
{f}_S \equiv {S \over S+B}   
       = {N_1 / N - \epsilon_B \over \epsilon_S - \epsilon_B } 
\eeq
where the total number of events is $N = N_1 + N_2 = S + B$. 
For any event-level binary tagging algorithm that is better than random with 
$\epsilon_S \neq \epsilon_B$ a measurement of $N_1, N_2$ yields a measurement 
of the fraction $f_S$. 
Note that for a perfect tagging algorithm with $\epsilon_S , \epsilon_B =1,0$ then 
$f_S = N_1 / N$. 
Note also that up to statistical and systematic uncertainties, 
the inferred fraction $f_S$ obtained from $N_1, N_2$ is independent of the tagging algorithm 
working point $\epsilon_S , \epsilon_B$ obtained for a given value of the classifier output variable cut or threshold.
This feature gives a simple internal 
consistency check of the tagging algorithm and measurement procedure, and to the extent that it does not hold, 
could provide a handle on systematic effects 
and uncertainties.  

A useful figure of merit to assess the utility of a binary tagging algorithm 
in a measurement as described above, is the 
statistical uncertainty that can be obtained for the fraction $f_S$. 
Neglecting any issues with systematic uncertainties,
the measured event counts in the two tagged categories $N_1,N_2$ may be treated as 
independent.   
In the limit $N_1, N_2 \gg 1$, the statistical uncertainty in the inferred fraction 
of events of one type is then given by the quadrature sum of the contributions from  
Poisson statistical uncertainties in each category 
\beq
\sigma_{\rm stat} ( f_S ) \simeq { 1 \over | \epsilon_S - \epsilon_B | } \, 
    \sqrt{  N_1 \,  N_2   \over N^3} 
    \label{fstat}
\eeq
For any event-level binary tagging algorithm that is better than random with 
$\epsilon_S \neq \epsilon_B$, the statistical uncertainty in $f_S$ can always 
be made small with a large enough sample of events. 
Of course in an actual measurement, systematic uncertainties often play an important 
role and limit the ultimate achievable uncertainty.


\subsection{Strange-Quark Jet Tagging in $W$-Boson Decays} 

The tagging algorithms studied in this paper are by construction only designed to discriminate strange- from down-quark jets. While this restriction is of course an idealization,  there  is at least one physical process that does yield mainly strange- and down-quark jets, namely the hadronic 
decay of a $W$-boson that includes a charm-quark jet. 
The $W$-boson can decay to a charm-quark and either a strange- down- or bottom-quark, 
$W \to c\, s  \, , \, c \, d \, , \, c \, b$. 
The strange- to down-quark ratio is large in these decays 
$ {\rm Br}( W \to c \, s) \, /  \, {\rm Br}(W \to c \, d) \simeq |V_{cs} / V_{cd} |^2 \simeq 20.9$
as is the down- to bottom-quark ratio 
$ {\rm Br}( W \to c \, d) \, /  \, {\rm Br}(W \to c \, b) \simeq |V_{cd} / V_{cb} |^2 \simeq 26.9$. 
So decays of a $W$-boson that include a charm-quark
represent a very high purity source of strange-quark jets, with a small fraction of down-quark jets, 
and an even smaller fraction of bottom-quark jets.

\begin{figure}[!t]
  \begin{center}
  \vspace{0.5cm}
   \includegraphics[clip, width=8cm]{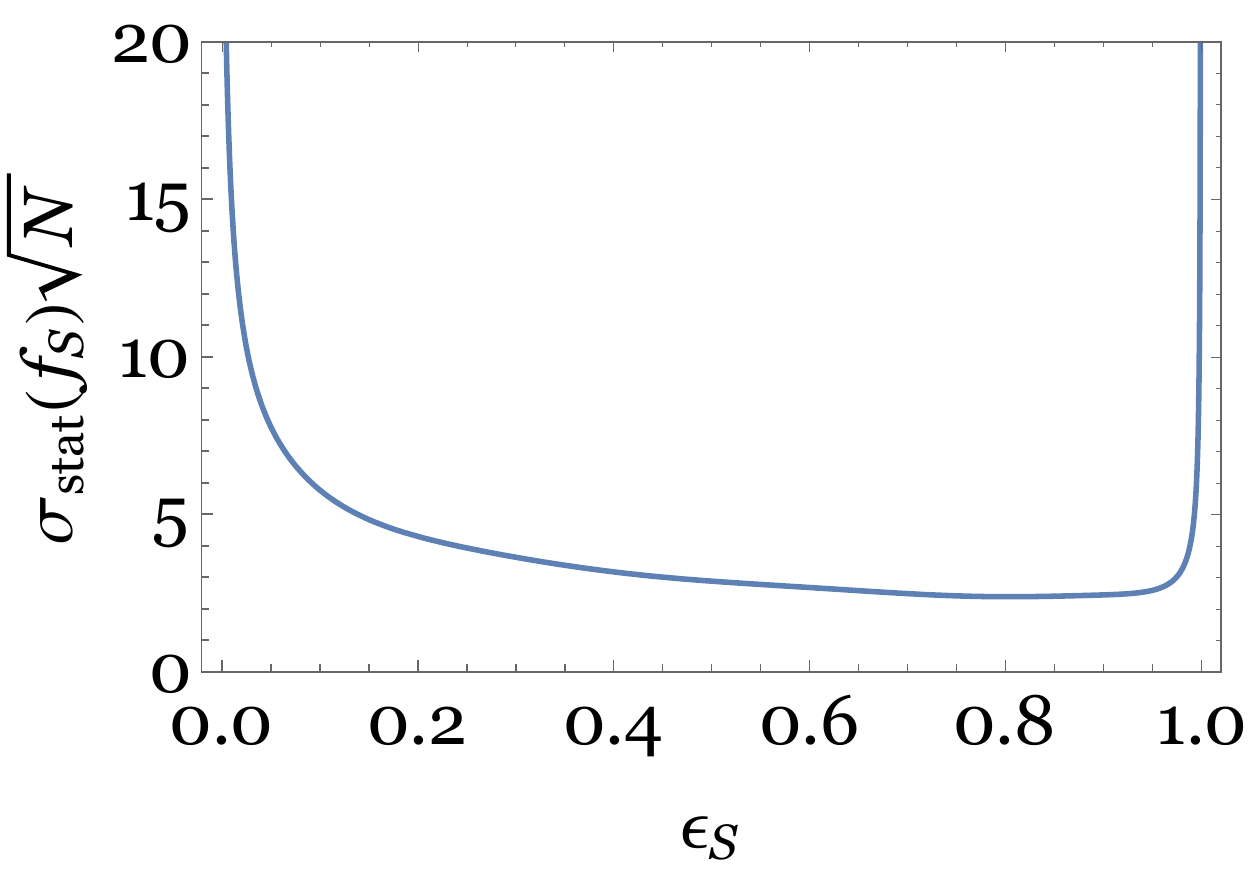}
  \end{center}
  \caption{The statistical uncertainty in the inferred fraction of strange-quark jets 
      times the square root of the total number of events 
    for $f_S \simeq 0.95$ 
    as a function of the strange-quark jet tagging efficiency 
      for  the CNN4 algorithm with $p_T = 45 $ GeV strange- and down-quark jets. 
}
  \label{fig:CKMratio}
\end{figure}

It is worthwhile to consider the 
performance of the binary tagging algorithms presented in this paper 
in a hypothetical background free 
measurement of the ratio of strange- to down-quark jets from a 
single isolated $W$-boson that decays hadronically to two resolved jets, 
where one of the jets is perfectly tagged as a charm-quark jet.  
While this is a highly idealized thought experiment that can not be achieved in reality 
in a high energy collider, it is none-the-less useful to illustrate the performance of 
strange- versus down-quark tagging algorithms.  
Neglecting the very small fraction of bottom-quark jets, 
the fraction of strange-quark jets in these decays is 
$f_S \simeq |V_{cs} |^2  / (  |V_{cs} |^2   +    |V_{cd} |^2 ) \simeq 0.95$. 
The statistical uncertainty figure of merit (\ref{fstat}) discussed in the previous 
sub-section, times the square root of the total 
number of events, all for this numerical value of the strange-quark jet fraction, 
is shown in 
Fig.~\ref{fig:CKMratio} as a function of the strange-quark jet tagging efficiency 
for the CNN4 algorithm with $p_T = 45 $ GeV jets.  
In this hypothetical measurement, the best statistical precision in the inferred 
strange-quark jet fraction with this algorithm occurs for strange-quark jet tagging 
efficiencies roughly in the range $0.2-0.9$. 
Near the upper end of this range where the statistical precision is best, 
the reduced discriminating power of the tagging algorithm is overcome by increased 
statistics.

At the LHC hadronic decays of $W$-bosons are most easily isolated from backgrounds 
in production of a top-quark and anti-top quark with 
semi-leptonic decays of the resulting $W$-bosons, 
$ t  \bar{t} \to Wb \, Wb \to \ell \nu b \, jj b $. 
Half of the time one of the two non-bottom-quark resolved jets in this final state is a charm-quark jet
(ignoring initial and final state radiation and jet overlaps).  
In this case, $ t  \bar{t} \to Wb \, Wb \to \ell \nu b \, j c \, b $, a fraction $f_S \simeq 0.95$ of the time 
the remaining resolved non-charm/bottom jet is a strange-quark jet, and a fraction 
$1 - f_S \simeq 0.05$ of the time it is a down-quark jet (again ignoring 
initial and final state radiation and jet overlaps). 
These particular final states could be largely isolated from backgrounds
and combinatoric confusion by requiring an isolated lepton, 
missing energy, at least four jets, two of which are bottom-quark tagged and one of which is charm-quark 
tagged along with kinematic constraints consistent with production and decay of a top-quark and anti-top quark 
and subsequent semi-leptonic cascade decays through $W$-bosons. 
Although to our knowledge charm-quark jet tagging has not been explicitly implemented in
resolved hadronic top-quark reconstruction, it could be included. 
So the process $ t  \bar{t} \to Wb \, Wb \to \ell \nu b \, j c \, b $
along with the resolved top-quark reconstruction outlined above, could provide 
a final state with a jet that may be isolated kinematically and which a large fraction 
of the time is a strange-quark jet. 
This could provide an interesting 
arena in which to apply and validate strange-quark jet tagging at the LHC.
Deployed in this way, the identification of strange-quark jets 
would have to be part of an all-jet tagging procedure that 
discriminated among all jet types. 
In this application this includes not only 
bottom- and charm-quark jets with displaced charged track 
vertices, but also other light-flavor and gluon jets from initial and final state radiation.


\section{Discussion \label{sec:discuss}}

Strange-quark tagging is the last largely unexplored problem in jet identification. 
In this paper we have demonstrated that
jets initiated by strange- and down-quarks 
can on average be distinguished by utilizing observables available to the current 
generation of general purpose high energy collider detectors, such as CMS and ATLAS 
at the LHC. 
We presented strange- versus down-quark jet tagging algorithms based on 
single whole-jet variables, BDTs of a few whole-jet variables, and 
CNNs with jet images based on inputs from detector sub-systems.  
These algorithms rely primarily on the observation that strange-quark jets have 
a higher momentum weighted fraction of neutral kaons that do down-quark jets. 
This difference can be observed and exploited in a general purpose collider detector in two main 
ways. 
First, long-lived neutral kaons, $K_L$, that are more common in strange-quark jets, 
deposit energy primarily in the HCAL, while neutral pions that decay promptly to 
photon pairs, $\pi^0 \to \gamma \gamma$, that are more common in 
down-quark jets deposit energy in the ECAL. 
So on average, strange-quark jets have a higher fraction of neutral hadronic energy, while 
down-quark jets have a higher fraction of neutral electromagnetic energy. 
Second, short-lived neutral kaons that decay-in-flight to charged pion pairs
within the inner tracking region, $K_S \to \pi^+ \pi^-$, can be identified. 
The reconstructed momentum fraction of these short-lived kaons is larger
for strange-quark jets than for down-quark jets. 

Strange-quark jet tagging, deployed as
an element of an all-jet tagging procedure that 
discriminated among all jet types, could have applications to many new physics searches 
and measurements  at high energy colliders. 
Forming event categories based on tagging algorithm output classifier variables 
can help to isolate sought out signals from backgrounds, always improves 
statistical uncertainties, and could even 
make new types of reconstruction and measurements possible. 
This could be especially beneficial in signatures with multiple jets. 
In particular, strange-quark jet tagging, 
in conjunction with bottom- and charm-quark tagging with displaced vertices,  
could help 
reduce combinatoric confusion in the reconstruction of resolved hadronic top-quark decays. 
For semi-leptonic top-quark anti-top quark signatures, 
hadronic top-quark reconstruction might be an interesting sub-process 
in which to attempt to demonstrate and validate 
strange-quark jet tagging at the LHC. 
Ultimately this might also make possible a direct measurement of the 
ratio of CKM matrix elements $| V_{cs} / V_{cd}|$ in hadronic $W$-boson decays 
within reconstructed hadronic top-quark decays. 
If possible, such a measurement would likely be limited by systematic rather than 
statistical uncertainties, but would be complementary to existing measurements 
of these individual 
CKM elements in meson decays that rely on hadronic form factors and lattice 
simulations \cite{Tanabashi:2018oca,Koppenburg:2017mad}. 

The focus of this work has been a demonstration of strange- versus down-quark jets 
using observables available in the current generation of general purpose high energy 
collider detectors.  
However, some special purpose detectors have detector elements that could 
provide additional observable handles to exploit in strange-quark jet tagging. 
A Cherenkov system, such as the one in the LHCb detector at the LHC, 
allows for charged particle discrimination, in particular between low to moderate momentum 
charged pions and charged kaons. 
While the momentum weighted sum of these two types of particles are very similar 
in strange- and down-quark jets, the individual momentum weighted fractions differ 
significantly, as can be seen clearly in Fig.~1. 
So charged particle identification with a Cherenkov system 
could provide further qualitative improvement over the 
performance of strange-quark 
jet tagging demonstrated here. 
Another possibility for achieving charged particle identification  is with 
precise timing information for charged particles. 
This can be used to deduce a velocity, that in combination with the standard 
measurement of momentum from track curvature in the magnetic field, yields 
a measure of the charged particle mass. 
The timing capabilities planned for future upgrades to the CMS and ATLAS detectors 
should allow separation of charged pions and kaons up to a few GeV in momentum. 
Timing resolutions sometimes discussed for 
future detectors at future colliders could increase this separation out to much larger momenta. 
This additional handle on the charged particle content of jets could improve many measured features  
of jet physics, including strange-quark jet tagging.


\section*{Acknowledgements}

We would like to thank Pouya Asadi, John Paul Chou, Anthony DiFranzo, Yuri Gershtein, 
Eva Halkiadakis, Amit Lath, Sebastian Macaluso, Angelo Monteux, and 
Sunil Somalwar for discussions.
This work was supported by the US Department of Energy under grant DE- SC0010008. D.S. is also supported in part by the Director, Office of Science, Office of High Energy Physics of the U.S. Department of Energy under the Contract No. DE-AC02-05CH11231. DS thanks LBNL, BCTP and BCCP for their generous support and hospitality during his sabbatical year. 
Y.N. is grateful to KEK for its hospitality during his stay when the paper was completed.
S.T. would like to thank the Kavli Institute for Theoretical Physics for its hospitality 
during completion of this work, made possible through support from the 
National Science Foundation under Grant NSF PHY-1748958.

\bibliography{ref}
\bibliographystyle{utphys}

\newpage

\appendix

\end{document}